\documentclass[%
 reprint,
 amsmath,amssymb,
 aps,
pre,
]{revtex4-2}

\usepackage{graphicx}
\usepackage{dcolumn}
\usepackage{bm}

\newcommand{\dbar}{d\hspace*{-0.08em}\bar{}\hspace*{0.1em}}

\begin{document}

\preprint{APS/123-QED}

\title{Scaling regimes and fluctuations of observables in computer glasses approaching the unjamming transition}

\author{Julia A. Giannini}
\affiliation{Department of Physics, Syracuse University, Syracuse, New York 13244, USA $\:$}%

\author{Edan Lerner}
\affiliation{Institute for Theoretical Physics, University of Amsterdam, Science Park 904, Amsterdam 1098 XH, Netherlands}%

\author{Francesco Zamponi}
\affiliation{Dipartimento di Fisica, Sapienza Universit\`a di Roma, Piazzale Aldo Moro 2, 00185 Rome, Italy}
\altaffiliation{Laboratoire de Physique de l'Ecole Normale Sup\'erieure, ENS, Universit\'e PSL, CNRS, Sorbonne Universit\'e, Universit\'e Paris-Diderot, Sorbonne Paris Cit\'e, Paris, France}%

\author{M. Lisa Manning}
    \email{mmanning@syr.edu}
\affiliation{Department of Physics, Syracuse University, Syracuse, New York 13244, USA}%

\date{\today}

\begin{abstract}
Under decompression, disordered solids undergo an unjamming transition where they become under-coordinated and lose their structural rigidity. The mechanical and vibrational properties of these materials have been an object of theoretical, numerical, and experimental research for decades. In the study of low-coordination solids, understanding the behavior and physical interpretation of observables that diverge near the transition is of particular importance. Several such quantities are length scales ($\xi$ or $l$) that characterize the size of excitations, the decay of spatial correlations, the response to perturbations, or the effect of physical constraints in the boundary or bulk of the material. Additionally, the spatial and sample-to-sample fluctuations of macroscopic observables such as contact statistics or elastic moduli diverge approaching unjamming. Here, we discuss important connections between all of these quantities, and present numerical results that characterize the scaling properties of sample-to-sample contact and shear modulus fluctuations in ensembles of low-coordination disordered sphere packings and spring networks. Overall, we highlight three distinct scaling regimes and two crossovers in the disorder quantifiers $\chi_z$ and $\chi_\mu$ as functions of system size $N$ and proximity to unjamming $\delta z$. As we discuss, $\chi_X$ relates to the standard deviation $\sigma_X$ of the sample-to-sample distribution of the quantity $X$ (e.g. excess coordination $\delta z$ or shear modulus $\mu$) for an ensemble of systems. Importantly, $\chi_\mu$ has been linked to experimentally accessible quantities that pertain to sound attenuation and the density of vibrational states in glasses. We investigate similarities and differences in the behaviors of $\chi_z$ and $\chi_\mu$ near the transition and discuss the implications of our findings on current literature, unifying findings in previous studies.
\end{abstract}


\maketitle


\section{\label{sec:intro}Introduction}

Jamming is a common phenomenon in nature - from clogging of grain hoppers, to traffic congestion and crowding in collective animal behavior, and biological tissue rigidification, objects collected at high enough densities behave as solids, resisting deformation and flow \cite{behringer_physics_2018, liu_jamming_1998, peter_crossover_2018, nagatani_physics_2002, lawson-keister_jamming_2021, garcimartin_flow_2015}. What is the smallest density that a collection of macroscopic objects could have such that it would still behave as a solid? What are the mechanical features of such a material? The physics of the jamming/unjamming transition seeks to answer these questions \cite{liu_jamming_2010, hecke_jamming_2009}.

In disordered systems of soft spheres under decompression, the unjamming transition occurs upon vanishing pressure $p$, where the material loses its ability to resist applied deformation and constituent particles begin to be able to move past each other freely. Understanding both i) the nature of this transition and ii) the mechanical and vibrational properties of materials near the transition is important for the development of predictive theories and for enabling functional material design. A number of theoretical, numerical, and experimental approaches have made progress toward this goal in recent years. Mean field theories connected the physics of spin glasses and unjamming phenomena, highlighting the complexity of the glassy potential energy landscape~\cite{parisi2020theory,muller2015marginal} and the existence of spatially extended excitations in  low-coordination solids~\cite{charbonneau_jamming_2015,franz2015universal}. Numerical and experimental studies formulated a jamming phase diagram (consisting of packing fraction $\phi$, shear stress $\tau$, and temperature $T$) and investigated the yielding behavior of deformed jammed solids \cite{behringer_physics_2018, liu_jamming_2010}.

Recent work studying a variety of modeled disordered materials has identified many length scales (denoted by $\xi$ or $l$) and frequency scales (denoted by $\omega$) that diverge or vanish respectively via power laws with decreasing excess coordination number, $\delta z \equiv z-z_c$ (where $z_c \equiv 2 d$). See Ref.~\onlinecite{lerner_quasilocalized_2018} for a comprehensive literature review \cite{karimi_elasticity_2015, baumgarten_nonlocal_2017, during_length_2014, bouzid_nonlocal_2013, wyart_geometric_2005}. Thus, theoretical work and simulation studies have focused on the critical behavior that occurs at the unjamming transition as well as the possible universal behaviors of low-coordination solids \cite{ikeda_dynamic_2013, charbonneau_jamming_2015}. A length scale of particular importance is the correlation length $\xi \sim \delta z^{-1/2}$ which has a variety of physical interpretations. First, it is associated with the breakdown of continuum elasticity, as the inherent structural disorder of a jammed solid becomes vitally important to its mechanical response \cite{lerner_breakdown_2014, mizuno_spatial_2016}. Consistent with this interpretation, $\xi$ represents the core size of low-energy excitations that constitute defects in glasses \cite{lerner_low-energy_2021}. Thus, several independent studies~\cite{karimi_elasticity_2015, baumgarten_nonlocal_2017, silbert_vibrations_2005, wyart_scaling_2010, ikeda_dynamic_2013, schoenholz_stability_2013, lerner_breakdown_2014, mizuno_spatial_2016, cakir_emergence_2016} measured $\xi$ numerically through examining the response of jammed disordered solids to multiple types of local, global, and boundary perturbations (see Ref.~\onlinecite{lerner_quasilocalized_2018}). Additionally, the scaling relationship $\xi \sim \delta z^{-1/2}$ was rationalized in Ref.~\onlinecite{Rens2019} and predicted via Effective Medium Theory (EMT) in Refs.~\onlinecite{wyart_scaling_2010,degiuli2014effects}, where $\xi$ corresponds to the transverse wavelength of vibrational modes in the boson peak \cite{silbert_vibrations_2005}.

In investigating the scaling properties and physical interpretation of diverging unjamming length scales, there are two important and understudied considerations that we highlight and characterize in this work. Although several existing studies have examined the significant finite size effects that occur in modeled jammed solids, few have determined how those effects might influence the measurement of $\xi$ or its relationship with $\delta z$. Particularly, given the relation $\xi \sim \delta z^{\gamma}$ where $\gamma = -1/2$ has been observed widely in the literature and predicted by EMT, it is possible that measurements of $\gamma$ that are close to, but distinct from  $\gamma = -1/2$ have been influenced by finite size effects\cite{goodrich_jamming_2014}. Further, it is not clear in the current literature how long power-law scaling relations such as $\xi \sim \delta z^{\gamma}$, and importantly, the underlying physics, persist in $\delta z$. As disordered solids become denser and more deeply quenched far from the unjamming transition, one expects that the length scale $\xi$ becomes irrelevant or exhibits a different scaling relationship with $\delta z$. Here, we clarify both of these concerns with numerical evidence and scaling arguments that are relevant to the correlation length $\xi$.

As in standard critical phenomena, the existence of a divergent length scale at unjamming implies that macroscopic quantities do not concentrate around their mean and show anomalous fluctuations. In generic disordered systems, fluctuations are due both to thermal effects within a single sample, and to the disorder that induces sample-to-sample fluctuations. Given that jammed packings are athermal, however, thermal fluctuations can be neglected and we thus focus here on sample-to-sample fluctuations. Indeed, work by Lerner et al. (Refs.~\onlinecite{kapteijns_unified_2021, gonzalez-lopez_variability_2023, gonzalez-lopez_mechanical_2021, lerner_anomalous_2023} and others), examined the relationship between sample-to-sample fluctuations of the shear modulus and several important vibrational ($A_g$ and $\Gamma(\omega)$, discussed in Sec.~\ref{sec:disorder_quant}  below) and mechanical ($\xi$) observables in ensembles of packings and spring networks.
For a given macroscopic observable $X$ (e.g. the shear modulus $\mu$ or coordination number $z$), one can introduce the normalized sample-to-sample standard deviation $\chi_X$ and study its behavior approaching unjamming. Because studies that sought to connect the underlying elasticity of low-coordination harmonic sphere packings with unjamming physics showed that, on average, the shear modulus $\mu$ scales linearly with $\delta z$ near the transition~\cite{matthieu_thesis}, a reasonable guess (that we will challenge below) is that the fluctuations of these quantities also scale together, i.e. $\chi_\mu \sim \chi_z$ \cite{goodrich_jamming_2014}.

More precisely, in Ref.~\onlinecite{gonzalez-lopez_variability_2023}, the scaling relation that the authors measured in spring networks between the so-called sample-to-sample disorder quantifier $\chi_\mu$ and $\delta z$ strongly supports a related scaling between the unjamming lengthscale $\xi$ and $\delta z$, $\chi_{\mu} \sim \xi \sim \delta z^{\gamma}$ with $\gamma = -1/2$ (we provide support for $\chi_{\mu} \sim \xi$ in Sec.~\ref{sec:regime3} below). Similar evidence for $\chi_{\mu} \sim \delta z^{-1/2}$ is also put forward in Ref.~\onlinecite{kapteijns_unified_2021}. Distinctly, in Ref.~\onlinecite{hexner_can_2019}, Hexner et al.~investigated both sample-to-sample and sub-system fluctuations of the contact number $z$ in harmonic packings and mean-field models. Ultimately, using the sample-to-sample disorder quantifier associated with contact number, $\chi_z$, the authors present numerical data that disagrees with $\gamma = -1/2$ in favor of a value numerically close to $-3/8$.
This discrepancy could have a variety of interpretations including i) the existence of another unjamming lengthscale, and ii) that there are subtleties in analysis ($\chi_{\mu}$ vs. $\chi_z$) or material preparation (system size effects, differences in microscopic interactions, proximity to unjamming, etc.) that affect the measurement of the exponent $\gamma$. Still, to form a more complete understanding of the properties of low-coordination disordered solids, it is important to resolve this inconsistency. In the discussion that follows, we will refer to $\gamma$ as a generic scaling exponent in the relation $\xi \sim \chi_X \sim \delta z^{\gamma}$. As discussed below, we find that $\gamma$ takes on different values in different contexts and model systems.

Here, we have expanded upon the work in Ref.~\onlinecite{hexner_can_2019}, to study sample-to-sample fluctuations of both $\mu$ and $z$ in large ensembles of harmonic packings and disordered spring networks in a wide range of $\delta z$. In addition to convincingly reproducing similar scaling behavior found in Ref.~\onlinecite{hexner_can_2019} (with $\vert \gamma \vert < 1/2$), we highlight three distinct scaling regimes and two crossovers for the disorder quantifier $\chi_z$ in harmonic packings as a function of $\delta z$  and system size $N$. Further, our results below provide a stringent  test of the hypothesis $\chi_\mu\sim\chi_z$ and show that it does not hold in all ranges of average coordination. As we discuss, $\chi_\mu$ is a very important measure of fluctuations that has been connected to interesting physical observables. Thus, developing more precise characterizations of both $\chi_z$ and $\chi_\mu$ is of crucial importance.

Overall, this study brings attention to effects of finite size and proximity to unjamming that are vital to consider to unify theories for and observations of the interesting properties of low-coordination solids. The work we present here stimulates future work in this area, and relates to a number of scaling behaviors relevant to anomalous elasticity, sound attenuation in amorphous solids, and the glassy vibrational density of states \cite{wang_sound_2019, degiuli_effects_2014, kapteijns_elastic_2021, szamel_microscopic_2022}.

\section{\label{sec:methods}Methods}

In this work we examine sample-to-sample contact and shear modulus fluctuations in both harmonic sphere packings and diluted spring networks approaching the unjamming transition. The disorder quantifiers $\chi_X$ that are the focus of our analysis relate distributions of macroscopic observables from ensembles of configurations (and ensembles of sub-systems) to the vibrational density of states, sound attenuation, and other quantities of interest such as the correlation/unjamming lengthscale $\xi$ described above \cite{gonzalez-lopez_mechanical_2021, kapteijns_unified_2021, gonzalez-lopez_variability_2023}. This section describes our numerical models as well as the computation and interpretation of $\chi_X$. 

\subsection{\label{sec:glass_model} Computer glass model}

This section discusses the computer glass model (similar to that described in Ref.~\onlinecite{hexner_can_2019}) from which we generate our main results regarding sample-to-sample contact fluctuations in systems approaching unjamming. We first note that although (as is common in the literature) we refer to our jammed harmonic sphere packings as glasses, they are indeed athermal. Further, we emphasize that our work studies only the (athermal) jamming transition that occurs with vanishing pressure, not the glass transition which is distinct, occurs as a function of (finite) temperature, and is the focus of a breath of other studies.

Individual realizations of our computer glass model consist of $N$ pairwise interacting particles in $\dbar = 3$ spatial dimensions and periodic boundary conditions. The harmonic pairwise potential $u_{\text{harm}}(r_{ij})$ depends only on the distance $r_{ij}$ between particles $i$ and $j$. We use a 50:50 binary mixture of harmonic spheres with unit mass and a 1:1.4 ratio between the particle diameters to prevent crystallization. The units of length in the simulations are chosen such that the diameter of the small particle species is $2R_{\text{small}} = 1.0$. The harmonic potential $u_{\text{harm}}(r_{ij})$ is given by
\begin{equation}
    u_{\text{harm}}(r_{ij}) =
    \begin{cases}
        \frac{1}{2}k(r_{ij}-l_{0,ij})^2, \: \: r_{ij} \leq l_{0,ij}\\
        0, \: \: r_{ij} > l_{0,ij}.\,
    \end{cases}
    \label{eqn:pair_energy_harm}
\end{equation}
where the constant $k$ is set to unity and the bond rest length $l_{0,ij}$ is equal to the sum of the radii of the particles $i$ and $j$, $l_{0,ij} = R_i + R_j$. The total energy $U$ of the system is equal to the sum of $u_{\text{harm}}$ over all interacting particle pairs.

To examine the behavior of this model approaching the unjamming transition, we prepare ensembles of packings at 16 different target pressures $p$ between $p = 10^{-6}$ and $p = 10^{-1}$ using the FIRE minimization algorithm and a Berendsen barostat \cite{lerner_mechanical_2019, bitzek_structural_2006}. We first prepared initial states at the highest pressure ($p = 10^{-1}$) by minimizing the total energy $U$ from random initial conditions. Then, we decompress the packings to each intermediate pressure state by repetitively changing the target pressure of the simulations and re-minimizing the energy. Since finite size effects are relevant to our analysis, we study large ensembles of packings at a variety of system sizes. Table \ref{tab:table1} below summarizes the number of independent samples $N_{\text{ens}}$ prepared at each system size for all pressure states.

\begin{table}[h]
\caption{\label{tab:table1} Ensemble sizes $N_{\text{ens}}$ for each system size $N$ and all pressures $p \in \left[10^{-6}, 10^{-1} \right]$}
\begin{ruledtabular}
\begin{tabular}{lcr}
System size $N$ & Ensemble size $N_{\text{ens}}$ \\
\hline
1000 & 9500\\
4000 & 9500\\
8000 & 5000\\
32000 & 1400\\
64000 & 650\\
\end{tabular}
\end{ruledtabular}
\end{table}

\subsection{\label{sec:net_model} Spring network model}

Several existing studies that utilize $\chi_{\mu}$ as a quantifier of mechanical disorder in low-coordination solids have used spring networks as a model system due to their simplicity and similarity to biological networks \cite{lerner_anomalous_2023, gonzalez-lopez_variability_2023}. The diluted networks included in this study are composed of relaxed Hookean springs that connect unit masses. Each network is initialized by adopting the disordered structure of a soft-sphere glass in $\dbar = 3$. We achieve this by placing a node at the center of each particle and a relaxed spring between each pair of interacting particles in the parent glass (see Ref.~\onlinecite{lerner_characteristic_2018} for a description of the glass model, which differs from the harmonic sphere model in Sec.~\ref{sec:glass_model}). Since the networks produced by this protocol have high coordination and we wish to study their behavior as a function of decreasing excess contact number $\delta z$, we then dilute the networks to reach a range of desired mean connectivities $\delta z \in \left[10^{-4}, 10^{1} \right]$. The edge-dilution algorithm is described in Ref.~\onlinecite{gonzalez-lopez_variability_2023} and involves criteria that minimize the coordination fluctuations of the resulting networks. For this reason, we study the shear modulus fluctuations of the networks via $\chi_\mu$ and do not measure $\chi_z$. The data reported below are from ensembles of at least 3000 independent networks at each value of $\delta z$ and $N \in \left\{512, 2048, 8192, 32768 \right\}$.

\subsection{\label{sec:disorder_quant} Disorder quantifier}

The parameter that is the focus of this study is the dimensionless disorder quantifier $\chi_X$, computed for both sample-to-sample fluctuations in excess contact number $\delta z$ and shear modulus $\mu$. Our computation of the shear modulus in this study follows the formalism detailed in Ref.~\onlinecite{hentschel_athermal_2011}. For an arbitrary macroscopic observable $X$, $\chi_X$ is computed as
\begin{equation}
    \chi_X = \frac{\sqrt{N \langle \left(X - \langle X \rangle \right)^2 \rangle}}{\langle X \rangle} = \sqrt{N}\frac{\sigma_X}{\langle X \rangle},
    \label{eqn:chiX}
\end{equation} 
where $\langle \cdot \rangle$ denotes an ensemble average, and $\sigma_X$ is the standard deviation of the sample-to-sample distribution of $X$. 

Since one of the goals of this work is to unify the findings of several existing studies that have examined sample-to-sample and sub-system fluctuations in low-coordination solids, we note that Ref.~\onlinecite{hexner_can_2019} examines contact fluctuations via a similar variance,
\begin{equation}
    \delta^2Z(N) \equiv N\left[\langle z^2\rangle - \langle z\rangle^2 \right] = (\chi_z \langle\delta z\rangle)^2.
\end{equation}
Thus, in our analysis below, when we determine the exponent $\gamma$ in the relation $\chi_z \sim \delta z^\gamma$, it will be equivalent to measuring the exponent $\nu_f = 2(\gamma+1)$ in Ref.~\onlinecite{hexner_can_2019}. 

In our comparison of $\chi_z$ and $\chi_\mu$, it is useful to note that the two quantities, while similar, exhibit different finite size effects. Particularly, the shear modulus fluctuations in small computer glasses at low pressure can be large enough to make determining a clear trend (i.e.~power law scaling) in $\chi_\mu(\delta z)$ difficult \cite{kapteijns_elastic_2021, mizuno_spatial_2016}. In other words, the measurement of $\chi_\mu$ is sensitive to outliers and can be quite noisy as a function of $\delta z$ or other control parameters. To address this challenge, Ref.~\onlinecite{kapteijns_unified_2021} examined several different methods for computing $\chi_\mu$: the direct method via Eq.~\eqref{eqn:chiX}, an outlier exclusion method, and a median method via:
\begin{equation}
    \chi_{X, \text{med}} = \frac{\sqrt{N \: \text{median}\left( \left(X - \langle X \rangle \right)^2 \right)}}{\langle X \rangle}.
    \label{eqn:chimed}
\end{equation} 
\noindent Ultimately, the authors showed that all three methods yield similar results, where $\chi_{\mu, \text{med}}$ is simpler to compute than implementing a outlier exclusion protocol and, in comparison to the direct measurement $\chi_{\mu}$, exhibits more consistent (smoothly-varying and less noisy) behavior as a function of their control parameter. Our results for shear modulus fluctuations in packings (not spring networks) below and in Appendices \ref{app:chi_mu_packings} and \ref{app:model_sys} thus feature $\chi_{\mu, \text{med}}$.

\begin{figure}[ht]
\includegraphics{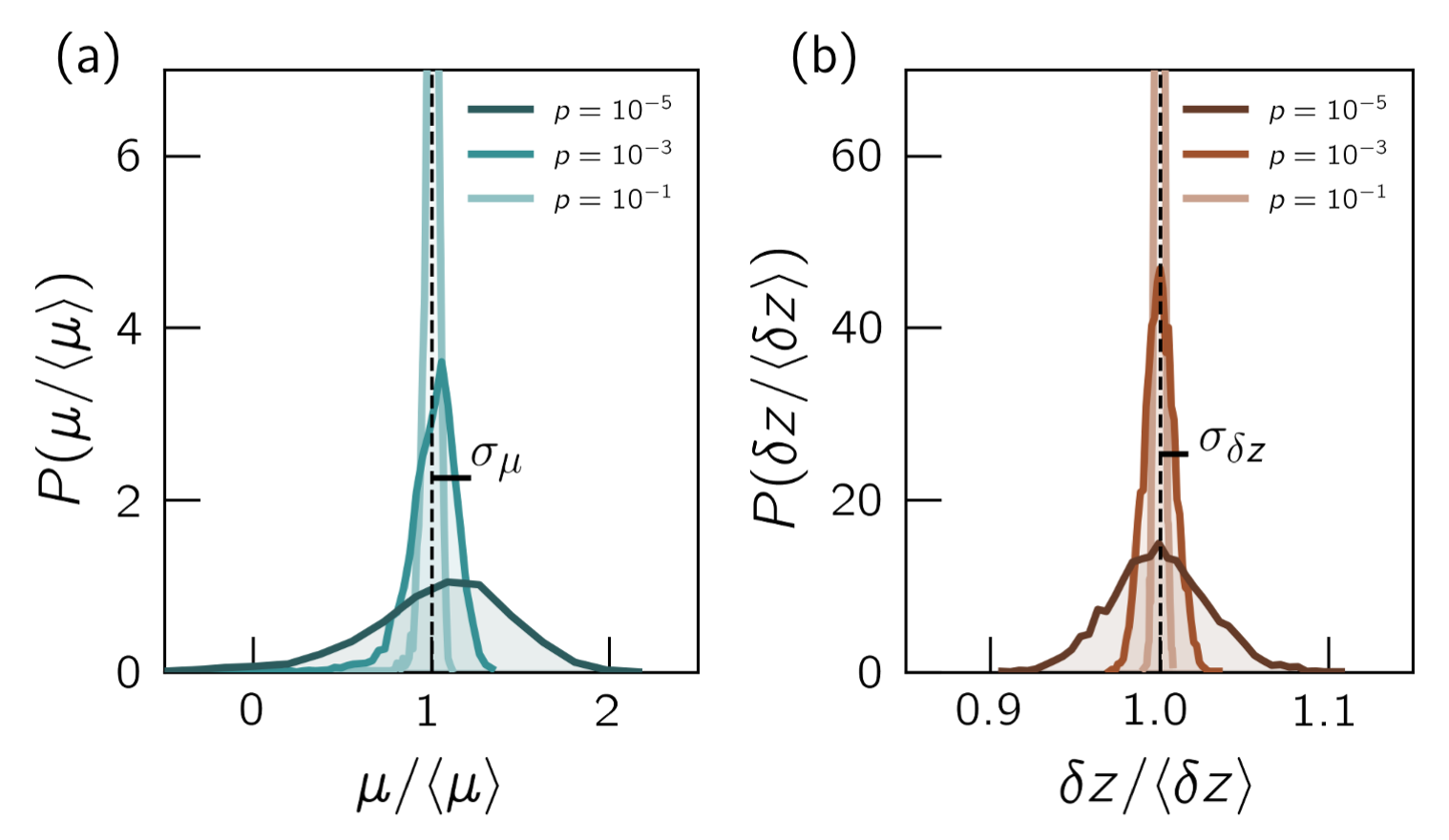}
\caption{\label{fig:hists} Sample-to-sample distributions of the shear modulus $\mu$ and excess contact number $\delta z$ rescaled by means for ensembles of harmonic sphere packings with $N = 8000$ and $p = 10^{-5},\: 10^{-3},$ and $10^{-1}$.}
\end{figure}

A key feature of disordered solids approaching the unjamming transition is the substantial broadening of the sample-to-sample distributions of $\mu$ and $\delta z$. This behavior is consistent with the divergence of the unjamming correlation length $\xi$ which we discuss further in the results that follow. Fig.~\ref{fig:hists} shows the distributions of $\mu$ and $\delta z$ for our ensembles of harmonic sphere packings with $N = 8000$ and $p = 10^{-5}, 10^{-3}, 10^{-1}$. $\chi_X$ is an effective parameter to capture this broadening and its scaling characteristics with $p$. While the distributions are qualitatively quite similar, the fluctuations of the shear modulus are $\sim 10$ times larger than those of the contact number.

The authors of
Refs.~\onlinecite{gonzalez-lopez_variability_2023, richard_brittle--ductile_2021, kapteijns_unified_2021} observed that $\chi_\mu$ is sensitive to control parameters involved in a wide variety of material preparation protocols: proximity to unjamming ($\delta z$), degree of annealing ($T_p$ as in Refs.~\onlinecite{gonzalez-lopez_variability_2023, kapteijns_unified_2021} or $\dot{T}$ as in Ref.~\onlinecite{richard_brittle--ductile_2021}), amount of internal stress ($\Delta$ as in Ref.~\onlinecite{gonzalez-lopez_variability_2023}), strength of long-range attraction between constituent particles ($Q$ as in Ref.~\onlinecite{gonzalez-lopez_variability_2023}), and others. Additionally, in numerical studies of amorphous plasticity and the brittle-to-ductile transition in  glasses under shear and tensile deformation, $\chi_{\mu}$ has been shown to relate to $A_g$, the density of micromechanical defects (or quasilocalized modes, QLMs) in computer glasses \cite{richard_brittle--ductile_2021, kapteijns_unified_2021}. Thus, $\chi_X$ is a useful parameter for characterizing changes in material behavior that result from general mechanical disorder.

Further motivating our investigation of $\chi_\mu$ as a physical observable, it has the distinct benefit of being experimentally accessible via its connection to the glassy vibrational density of states (vDOS), Fluctuating Elasticity Theory (FET), and wave attenuation rates \cite{ramos_heterogeneous_2022, schirmacher_comments_2011, schirmacher_acoustic_2007, kapteijns_unified_2021, schirmacher_thermal_2006}. Specifically, in Ref.~\onlinecite{kapteijns_unified_2021}, the authors study a \textit{distinct but related} disorder quantifier (also denoted $\chi$) that characterizes the spectral broadening of discrete phonon bands that occurs in the low-frequency vibrational spectra of computer glasses created by quenching systems equilibrated (via swap-Monte-Carlo) at varying `parent' temperatures $T_p$. They show that spectral broadening occurs with increasing $T_p$, which also corresponds to increasing mechanical disorder (quantified by $\chi$ or $\chi_{\mu}$). Further, assuming the equivalence (demonstrated in Ref.~\onlinecite{mahajan_unifying_2021}) of spatial fluctuations of the shear modulus over a coarse-graining lengthscale and sample-to-sample fluctuations ($\chi_{\mu}$ as we define it here), the same work\cite{kapteijns_unified_2021,gonzalez-lopez_mechanical_2021} supports the predictions of FET, that the low-frequency wave attenuation rate $\Gamma$ scales with $\chi_{\mu}^2$ and has frequency dependence $\omega^{\dbar +1}\:$. 

\begin{figure*}[ht]
\includegraphics{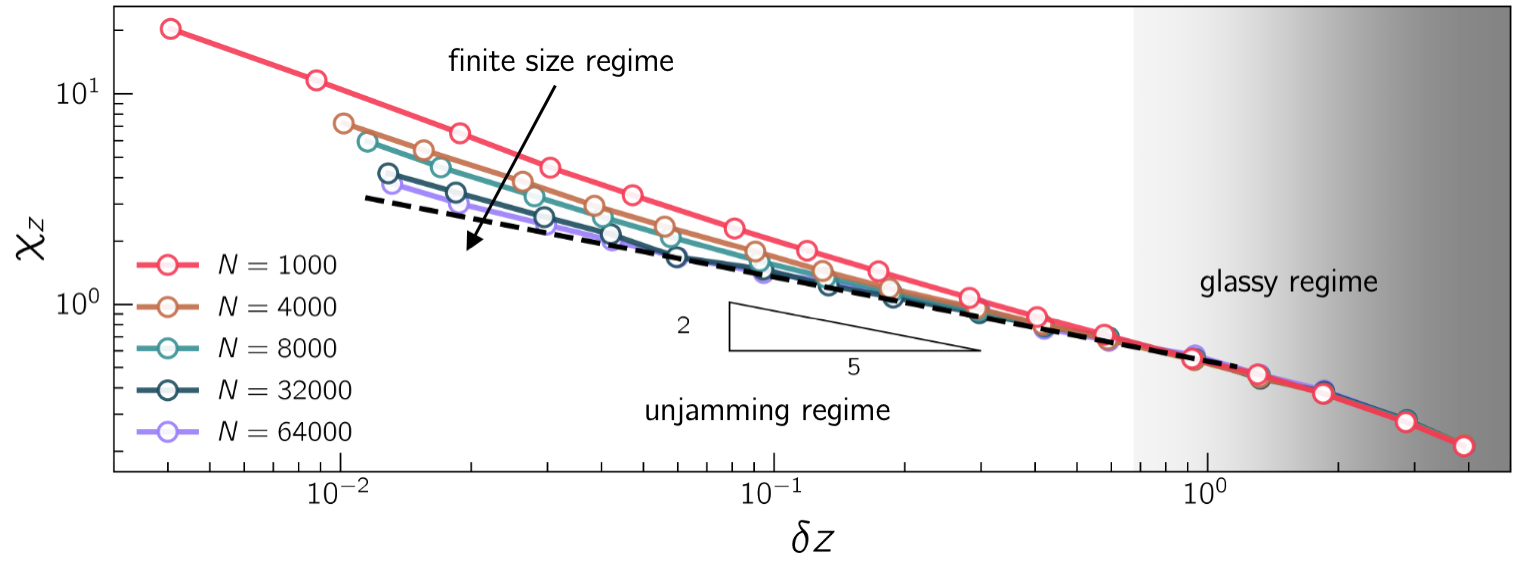}
\caption{\label{fig:chiz_regimes}Contact number disorder quantifier $\chi_z$ as a function of $\delta z$ for all system sizes (increasing from top to bottom) of harmonic sphere packings in $\dbar = 3$. Three distinct scaling regimes are visible as discussed in the main text. The dashed line is a guide to the eye and has a slope of $\gamma \approx -2/5$. The shaded region depicts the $N$-independent glassy regime for $\delta z \lesssim 7 \times 10^{-1}$.} 
\end{figure*} 

Now, in the discussion of the unjamming/QLM core-size lengthscale $\xi$ that follows in Sec.~\ref{sec:regime3}, we will relate the square of the disorder quantifier to a correlation volume. This is a reasonable assumption given the proposed equivalence between spatial and sample-to-sample fluctuations of the shear modulus, but remains to be demonstrated directly \cite{kapteijns_unified_2021,mahajan_unifying_2021}. Regarding this equivalence, note that in Ref.~\onlinecite{hexner_can_2019}, the authors show that spatial and sample-to-sample \textit{contact} fluctuations are not equivalent for finite $N$ and constant $\delta z$. Rather, the sub-system fluctuations are much larger than the sample-to-sample ones and grow much faster with decreasing $N$. However, it may be the case that contact and shear modulus fluctuations differ in this feature, or that we can proceed with assuming proportionality between sub-system and sample-to-sample fluctuations of the contact number at fixed $N$ and varying $\delta z$. The primary reasoning for examining sample-to-sample fluctuations over spatial fluctuations is computational simplicity. We also note that computing $\chi_X$ is generally simpler than obtaining direct measurements of $\xi$.

To summarize this section and highlight the utility of $\chi_X$ (with $X\rightarrow \mu, \delta z$) as a quantifier of mechanical disorder, we emphasize that $\chi_X$ is relevant to understanding the physical properties of low-coordination disordered solids in several related contexts. First, $\chi_X$ is sensitive to several model parameters that control the overall stability of glassy samples. Second, $\chi_\mu$ has been shown to relate to the density ($A_g$) and size ($\xi$) of glassy defects. Last, $\chi_\mu$ relates to experimentally accessible quantities via wave attenuation rates and the glassy vDOS. Overall, the inherent structural disorder and corresponding unique mechanical behaviors of glasses are directly reflected in the spatial and sample-to-sample fluctuations of $\mu$ and/or $\delta z$.

\section{\label{sec:results} Results}

As mentioned in Sec.~\ref{sec:intro}, by analyzing contact fluctuations in ensembles of harmonic packings, we have identified three scaling regions in $\chi_z(N, \delta z)$ that are relevant to understanding the relationship between structural disorder and material behavior in the proximity of the unjamming transition. We investigate the value of the scaling exponent $\gamma$ (in the relation $\chi_X \sim \delta z^\gamma$) in both packings and disordered spring networks and discuss the dependence of the results on model and analysis details. Additionally, by associating the scaling behavior of the disorder quantifier $\chi_X$ with that of the correlation lengthscale $\xi$, we discuss the crossover from unjamming-like behavior to that of higher coordination (more deeply annealed) glasses. This section presents our numerical results in the context of existing studies and theories. 

\subsection{\label{sec:scaling_regimes} Scaling regimes of contact fluctuations in packings}

To begin, we consider the scaling of $\chi_z$ with $\delta z$ in our ensembles of three-dimensional harmonic sphere packings under decompression. This analysis closely follows the results for sample-to-sample coordination fluctuations presented in Ref.~\onlinecite{hexner_can_2019}, although we explore a larger range of $\delta z$ here. Fig.~\ref{fig:chiz_regimes} highlights one of the main results of this work, that there are three distinct scaling regimes in $\chi_z$ as a function of $\delta z$ and $N$: (i) a finite size regime at low $\delta z \lesssim 5 \times 10^{-2}$ and small $N \lesssim 8000$, (ii) an unjamming regime at low-intermediate $\delta z \sim 10^{-1}$ and large $N \sim 32000$, and (iii) a system size independent so-called glassy regime at $\delta z \sim 7 \times 10^{-1}$. We proceeding by discussing the functional behavior in each regime separately. 

\subsubsection{\label{sec:regime1} Finite size regime}

As we show in Fig.~\ref{fig:chiz_regimes}, at low $\delta z \lesssim 5 \times 10^{-2}$ and small $N \lesssim 8000$, $\chi_z$ increases significantly with decreasing system size. Once the system size is sufficiently large, $\chi_z$ converges to a finite value at constant $\delta z$ (which corresponds to the unjamming regime described next). To characterize the dependence of $\chi_z$ on system size, we follow a similar scaling procedure to the one in Ref.~\onlinecite{hexner_can_2019}. First, we assume that $\chi_z$ scales with $N$ and $\delta z$ as
\begin{equation}
    \chi_z \sim N^{-\beta}f(\delta z\: N^\alpha),
\end{equation} 
where $\alpha$ and $\beta$ are to be determined below, and in the large-argument limit the function $f$ achieves $f(x \rightarrow \infty) \sim x^{\gamma}$. Because at large $\delta z$, $\chi_z$ is independent of $N$, we must have that $\beta = \alpha \gamma$. Thus, we write
\begin{equation}
    \chi_z \: N^{\alpha \gamma} \sim f(\delta z \: N^{\alpha}).
    \label{eqn:Nrescaling}
\end{equation}
Since this scaling form with $N$ depends on the determination of the exponent $\gamma$ associated with the unjamming regime, we now turn to the data collapses presented in Fig.~\ref{fig:chiz_collapse}. Note that the finite size effects associated with $\chi_z$ observed here are likely relevant in many numerical studies of low-coordination amorphous particulate solids. 
\subsubsection{\label{sec:regime2} Unjamming regime and scaling exponents}

The second scaling regime we identify in $\chi_z(N,\delta z)$ is the unjamming regime that occurs for $\delta z \lesssim 7 \times 10^{-1}$ in the limit $N \rightarrow \infty$ (see Fig.~\ref{fig:chiz_regimes}). For finite size systems, the scaling of $\chi_{z}$ crosses over from the finite size to the unjamming regime roughly around $\delta z \sim 10^{-1}$ (depending on $N$). In this range of $\delta z$, the contact fluctuations (quantified by $\chi_z$) scale as $\chi_z \sim \delta z^{\gamma}$. As we emphasized in Sec.~\ref{sec:intro}, the exponent $\gamma = -1/2$ was both theoretically predicted and numerically demonstrated in existing studies of disordered solids approaching unjamming \cite{wyart_scaling_2010, silbert_vibrations_2005}. This scaling is often associated with the correlation length $\xi$ or a corresponding correlation volume. However, in Ref.~\onlinecite{hexner_can_2019}, the authors demonstrated (via $\delta^2 Z$ as defined in Sec.~\ref{sec:disorder_quant}) that $\gamma \approx -3/8$ in harmonic sphere packings with $\dbar = 3$ and $4$ and mean field models with $\dbar = 2$ and $3$. While the values of these exponents are quite close, it is important to distinguish between them to understand the relationship between disorder and mechanics in low-coordination disordered solids. 

\begin{figure}[h]
\includegraphics{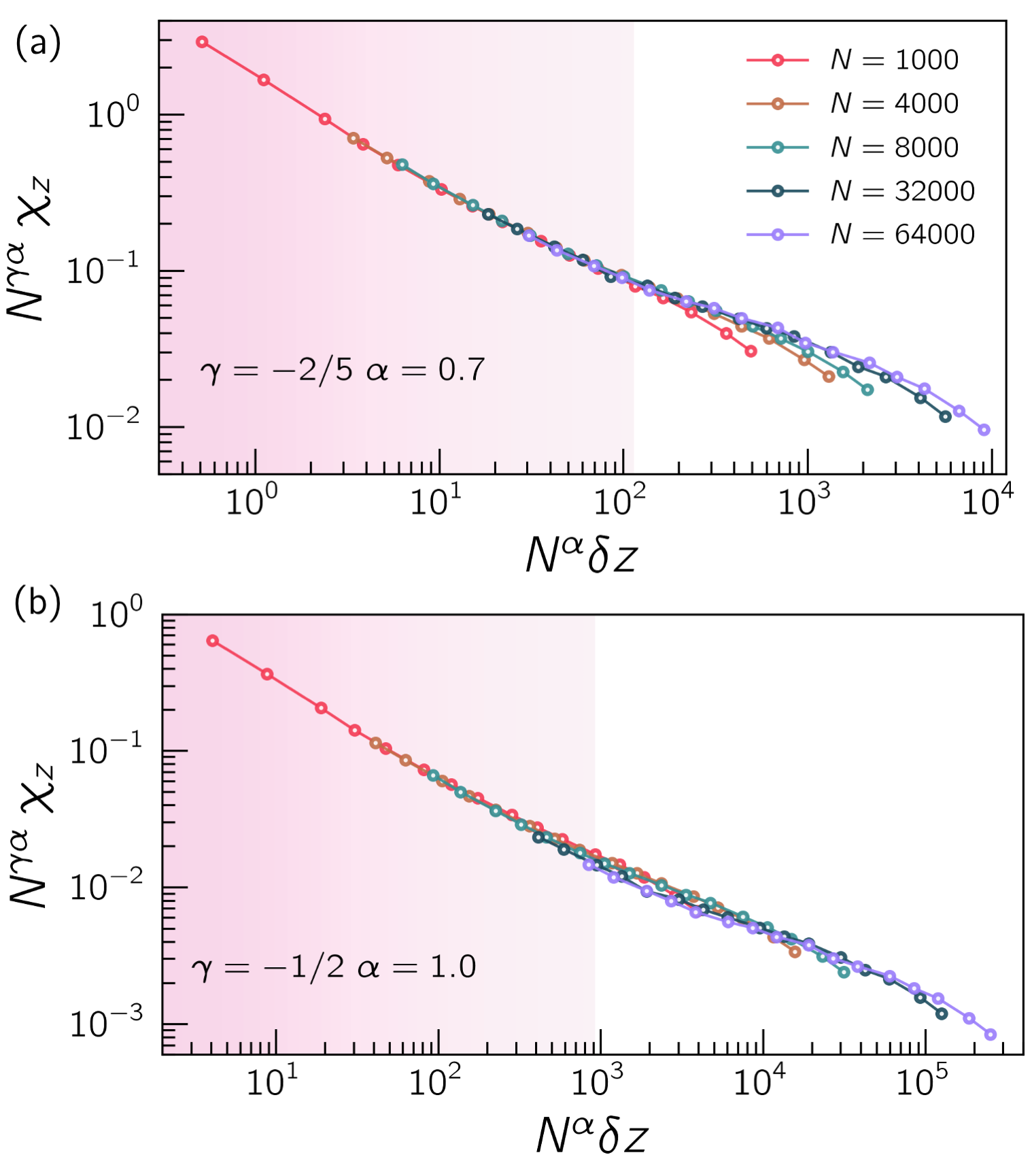}
\caption{\label{fig:chiz_collapse} Scaling collapses via Eq.~\eqref{eqn:Nrescaling} comparing values of $\gamma$ for the same data as shown in Fig.~\ref{fig:chiz_regimes}. The red shaded region depicts the collapse of both the finite size and unjamming regimes up to the onset of the glassy regime for the smallest system size $N = 1000$ (after which the data for that system size is not expected to collapse).}
\end{figure}

Here, we determine $\gamma$ via the scaling form proposed in Eq.~\eqref{eqn:Nrescaling}, by fitting the large-$N$ data in the unjamming regime to the power law $\chi_z \sim \delta z^{\gamma}$. Then, to determine the finite size behavior described above, we plot $\chi_z N^{\alpha \gamma}$ vs. $\delta z N^{\alpha}$ and alter $\alpha$ until the data collapses up to $\delta z \sim 7\times 10^{-1}$, where $\chi_z(N, \delta z)$ exhibits a second scaling crossover into the glassy regime. In this representation, we do not expect a data collapse in the regions corresponding to the high-coordination solids for each system size (where $N^{\alpha}\delta z$ exceeds $\sim N^{\alpha}(7 \times 10^{-1})$). Fig.~\ref{fig:chiz_collapse} summarizes the results of this scaling analysis, where panel (a) shows the collapse for $\gamma \approx -2/5$ and $\alpha = 0.7$ and (b) for $\gamma = -1/2$ and $\alpha = 1.0$. While both collapses are reasonably good, the first set of exponents, consistent with Ref.~\onlinecite{hexner_can_2019}, produce a better collapse in both the finite size and unjamming regimes. The data collapses in these two regimes are highlighted for example in Fig.~\ref{fig:chiz_collapse} up to the values of $N^{\alpha}\delta z$ where the $N=1000$ systems cross over to the glassy regime. An alternative representation of this finite size analysis is presented in Appendix \ref{app:chiz_reduced}. 

We note here that there is a slight difference between the exponent $\gamma \approx -2/5$ that we report here and the measurement of $\gamma \approx -3/8$ by Hexner et.~al.~However, $\gamma \approx -3/8$ and $\alpha = 0.6$ also work quite well for our $\chi_z(N,\delta z)$ data. Viewing the rescaling of the $\chi_z$ data in Appendix \ref{app:model_sys}, Fig.~\ref{fig:models_scaling} below, we favor $\gamma \approx -2/5$. Overall, the values of the scaling exponents here and in Ref.~\onlinecite{hexner_can_2019} are only approximate, and we emphasize that our analysis still supports the results of Ref.~\onlinecite{hexner_can_2019}, particularly the fact that $\vert \gamma \vert < 1/2$.

\subsubsection{\label{sec:regime3} Glassy regime and lengthscale crossover}

The final scaling regime that we identify in $\chi_z(N,\delta z)$ is the $N$-independent glassy regime, which occurs for $\delta z \gtrsim 7 \times 10^{-1}$ and is depicted in the shaded region of Fig.~\ref{fig:chiz_regimes}. In this region, it appears that the scaling of $\chi_z$ with $\delta z$ crosses over from $\chi_z \sim \delta z^{\gamma}$ to a steeper negative power-law. However, we cannot precisely determine a power-law scaling exponent for $\chi_z$ deep in the glassy regime, as our harmonic sphere packings become over-compressed and have trivial structure for $\delta z \gtrsim 10$ due to the absence of a strong short-range repulsion. To further investigate this point, future studies might examine a different glass model which produces more well-annealed (high coordination) samples. We note here that Ref.~\onlinecite{gonzalez-lopez_variability_2023} studies the behavior of the disorder quantifier $\chi_{\mu}$ in a wide variety of modeled disordered solids, but only briefly investigates the effect of varying the excess contact number $\delta z$. 

\begin{figure}[ht]
\includegraphics{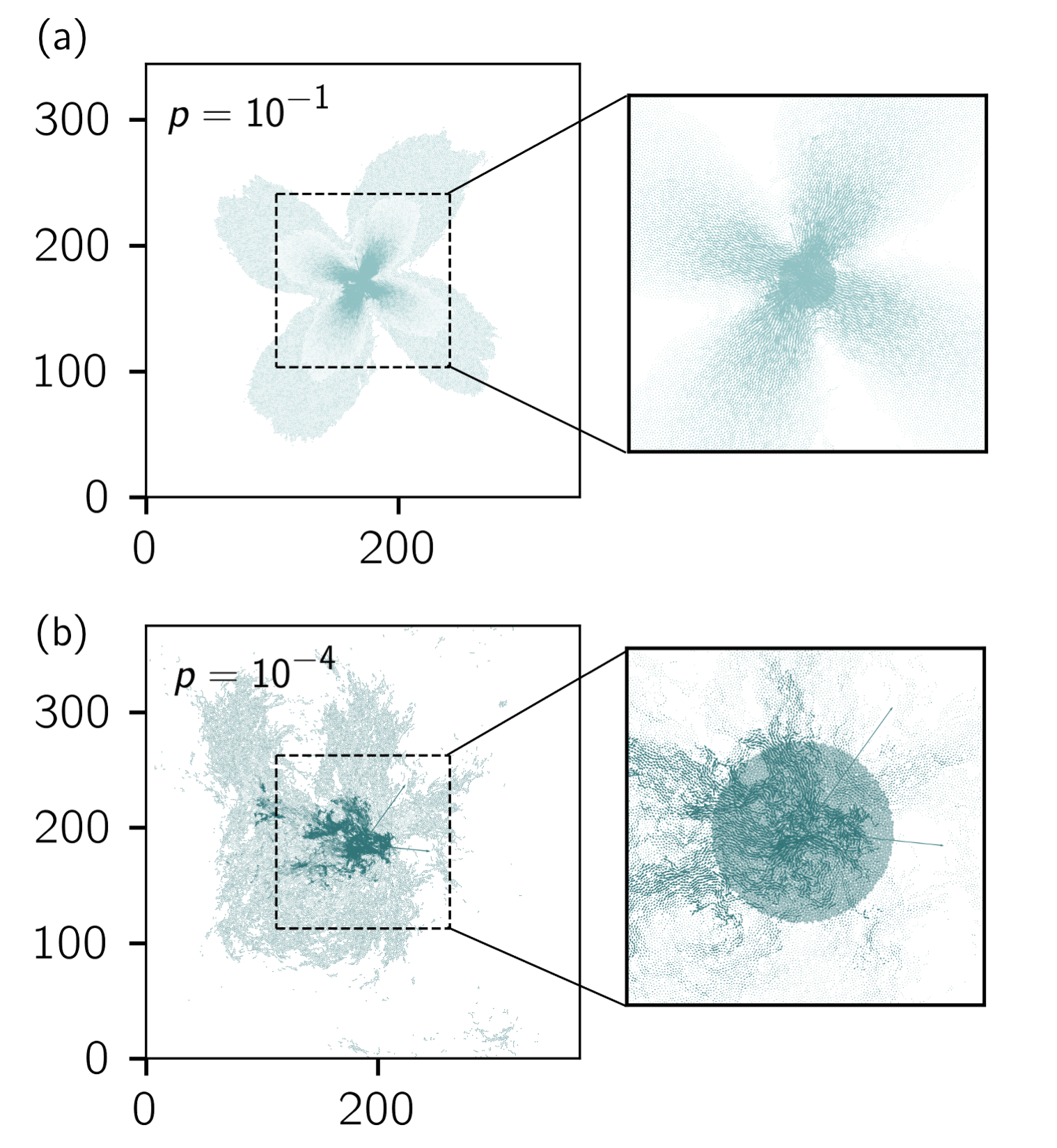}
\caption{\label{fig:dipole_res} Displacement response fields that result from dipolar forces applied to particle pairs in harmonic sphere packings with $\dbar = 2$ (for illustrative purposes) and $N = 102400$, at $p = 10^{-1}$ and $p = 10^{-4}$. We show only the displacement vectors with the top $25\%$ in magnitude. The insets show the approximate relative sizes of the disordered cores of the response fields, where the core grows substantially upon decreasing the pressure of the packings.}
\end{figure}

To explain the occurrence of the second scaling crossover in $\chi_z$ around $\delta z \sim 7 \times 10^{-1}$, we will draw upon the work of Ref.~\onlinecite{lerner_anomalous_2023}, where the authors studied a crossover in anomalous to elastic behavior in the displacement response $\vec{u}(\vec{r})$ of disordered spring networks to applied force dipoles. These displacement response fields have been studied thoroughly in modeled amorphous solids, particularly due to their structural similarity to quasilocalized vibrational modes (QLMs), which are good representations of microstructural defects in glasses \cite{lerner_low-energy_2021, rainone_pinching_2020, lerner_anomalous_2023, lerner_breakdown_2014, giannini_bond-space_2021}. Fig.~\ref{fig:dipole_res} depicts two such displacement fields for $\dbar = 2$ harmonic sphere packings with $p = 10^{-1}$ and $p = 10^{-4}$. We obtain the displacement fields through a similar numerical procedure to that described in Refs.~\onlinecite{lerner_breakdown_2014,rainone_pinching_2020}.

In Ref.~\onlinecite{lerner_anomalous_2023}, Lerner et.~al.~present numerical evidence for the following form of dipolar response fields in dimension $\dbar$, where the `core' behavior for $r \ll \xi$ is caused by micromechanical disorder and for $r \gg \xi$ is the result from continuum linear elasticity:
\begin{equation}
    |\vec{u}(\vec{r})| \sim \begin{cases}
  r^{-\frac{(\dbar - 2)}{2}},  & r \ll \xi \\
   r^{-(\dbar - 1)},  & r \gg \xi.
    \end{cases}
    \label{eqn:urdecay}
\end{equation}
This difference in behavior in $\vec{u}(\vec{r})$ is visible in Figs.~\ref{fig:dipole_res}a and \ref{fig:dipole_res}b. The packing depicted in Fig.~\ref{fig:dipole_res}a was prepared at a relatively high overall pressure ($p = 10^{-1}$), where continuum elasticity is relevant to the behavior on a macroscopic length scale. Thus, the dipole response field features a small disordered core decorated by a decaying quadrupolar field. In Fig.~\ref{fig:dipole_res}b, the packing was prepared at a pressure corresponding to the unjamming regime ($p = 10^{-4}$), and the behavior of the dipole response field is not as characteristic of continuum elasticity. Rather, the displacement field features a much larger core and a more disordered decaying elastic field. Recall from Sec.~\ref{sec:intro} above that the diverging length scale $\xi \sim \delta z^{-1/2}$ is associated with the core size of glassy defects. Since the dipolar response fields pictured above are structurally extremely similar to QLMs, we take Fig.~\ref{fig:dipole_res} and Fig.~\ref{fig:dipoles_all} in Appendix Ref.~\ref{app:dipole_responses} to somewhat directly depict the growth of $\xi$ approaching unjamming \cite{lerner_low-energy_2021, rainone_pinching_2020}.

From Eq.~\eqref{eqn:urdecay}, the authors of Ref.~\onlinecite{lerner_anomalous_2023} construct a correlation function $C(r) = \langle \vec{u}(\vec{r}) \cdot \vec{u}(\vec{r}) \rangle$ and a corresponding correlation volume $V_\xi$ via $V_\xi \sim \int_0^{\xi} d^{\dbar}r \; C(r)$. Thus, following Ref.~\onlinecite{lerner_anomalous_2023}, when we compute this correlation volume for $r\ll \xi$, we obtain $V_\xi \sim \xi^{2}$ independent of $\dbar$ (consistent with Refs.~\onlinecite{shimada_spatial_2018, yan_variational_2016, ji_mean-field_2022}). Since the elastic decay for $r \gg \xi$ in Eq.~\eqref{eqn:urdecay} is fast compared to the anomalous behavior for $r \ll \xi$, we next assume that for $r \gg \xi$, $V_\xi$ scales with $\xi$ in the `naive' way, corresponding to random fluctuations in material properties $V_{\xi} \sim \xi^{\dbar}$ (see Ref.~\onlinecite{lerner_anomalous_2023} for details). Finally, if we accept that $\chi_X^2 \sim V_\xi$ and that $\xi \sim \delta z^{-1/2}$,  we have: 
\begin{equation}
    \chi_X \sim \begin{cases}
  \xi \sim \delta z^{-1/2},  & \delta z \lesssim 10^{-1} \text{ and } \xi \gg \mathcal{O}(10) \\
   \xi^{\dbar/2} \sim \delta z^{-\dbar/4},  & \delta z \gtrsim 1 \text{ and } \xi \sim \mathcal{O}(10),
    \end{cases}
    \label{eqn:chi_crossover}
\end{equation}
where $\xi \sim {\cal O}(10)$ is the typical core size of QLMs in well-annealed glasses \cite{lerner_low-energy_2021, rainone_pinching_2020}. Noticeably, this scaling relationship for $\chi_z$ with $\delta z \lesssim 10^{-1}$ is \textit{not} consistent with the above obervation, that $\gamma \approx -2/5$. As we discuss below, this deviation can likely be attributed to the difference between contact and shear modulus fluctuations in our harmonic sphere packings.

While Eq.~\eqref{eqn:chi_crossover} involves several assumptions, the general intuition is that if low-coordination glassy samples have large anomalous cores in the field $\vec{u}(\vec{r})$, then $\chi_\mu$, which can be measured from local spatial fluctuations of elastic moduli, will reflect that behavior in its scaling with $\xi$ and $\delta z$. If we consider instead a high-coordination sample, the core of $\vec{u}(\vec{r})$ will be small, and local mechanical fluctuations will not pick up the anomalous behavior. 

Overall, the scaling behavior of $\chi_z(\delta z)$ in the glassy regime (Fig.~\ref{fig:chiz_regimes}) is consistent with the direction of the crossover predicted by Eq.~\eqref{eqn:chi_crossover}, although as mentioned above, a clean power-law scaling at high-$\delta z$ is not discernible from our dataset. Still, it is important to note that there is likely a simultaneous crossover in $\xi(\delta z)$ that occurs around $\delta z \sim 7 \times 10^{-1}$, which would cause the behavior of $\chi_z(\delta z)$ to further deviate from Eq.~\eqref{eqn:chi_crossover} \cite{gonzalez-lopez_mechanical_2021, gonzalez-lopez_variability_2023}. Further classifying the behavior of $\chi_X(\xi)$ and $\xi(\delta z)$ is an interesting avenue for future work. 

\subsection{\label{sec:springs}Disorder quantifier in spring networks}

Having examined the behavior of contact fluctuations in modeled glasses under decompression, we now study $\chi_\mu(N, \delta z)$ in diluted spring networks. Ref.~\onlinecite{gonzalez-lopez_variability_2023} recently presented a similar dataset, supporting $\gamma = -1/2$. As mentioned in Sec.~\ref{sec:disorder_quant} above (and demonstrated in Appendix \ref{app:chi_mu_packings} below), $\chi_{\mu}$ measured in our ensemble of harmonic packings does not exhibit clean scaling behavior compared to $\chi_{z}$ in packings, or as we will see, $\chi_{\mu}$ in networks.

\begin{figure}[h]
\includegraphics{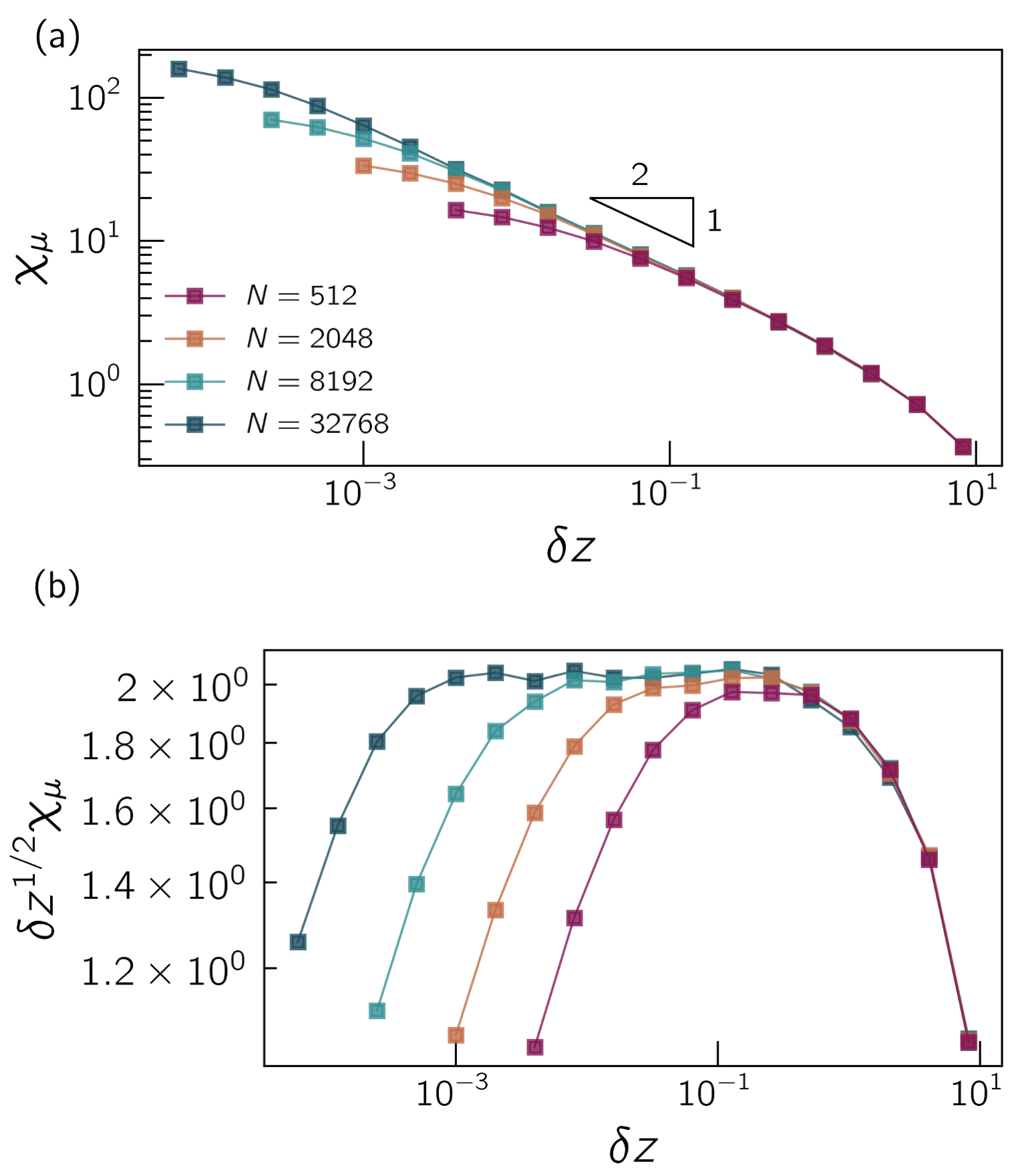}
\caption{\label{fig:springs_scaling} (a) Shear modulus disorder quantifier $\chi_\mu$ as a function of $\delta z$ for all system sizes (decreasing from top to bottom) measured in ensembles of diluted spring networks. (b) The same data as panel (a), rescaled by $\delta z^{1/2}$ on the vertical axis, to verify the relation $\chi_{\mu} \sim \delta z^{-1/2}$.}
\end{figure}

Fig.~\ref{fig:springs_scaling}a shows $\chi_\mu(\delta z)$ for ensembles of networks with several different system sizes $N$. Similarly to our analysis of $\chi_z$ above, we notice the same three scaling regimes and two crossovers as a function of $\delta z$: finite size, unjamming, and glassy. However, the finite size regime here exhibits the opposite behavior with increasing $N$ and constant $\delta z$ to that of $\chi_z$ in packings. That is, $\chi_{\mu}$ increases with $N$ to a constant value instead of decreasing to a constant value as $\chi_{z}$ does. Further, in the unjamming regime, the scaling of $\chi_\mu$ with $\delta z$ is quite convincingly characterized by $\gamma = -1/2$. This is further evidenced by the representation in Fig.~\ref{fig:springs_scaling}b, where we show $\chi_{\mu}/\delta z^{-1/2}$ vs. $\delta z$ and the region of $10^{-3} < \delta z \lesssim 10^{-1}$ is horizontal for large-$N$. Last, we note that the scaling crossover into the glassy regime around $\delta z \sim 7 \times 10^{-1}$ follows the general behavior predicted by Eq.~\eqref{eqn:chi_crossover}.

Clearly, there are distinct behaviors exhibited in $\chi_z(\delta z, N)$ measured in packings under decompression and $\chi_\mu(\delta z, N)$ measured in diluted spring networks in the finite size and unjamming regimes. One hypothesis to explain these differences is that packings under decompression undergo significant nonaffine rearrangements that lead to differences in local structure, while the diluted networks have minimal changes in coordination (see Sec.~\ref{sec:net_model} above). However, in Appendix \ref{app:model_sys} we compare $\chi_\mu(\delta z, N \sim 8000)$ in the diluted networks and spring networks derived directly from our harmonic sphere packings. We observe little-to-no difference in behavior, except that the crossover to the glassy regime occurs slightly quicker in the packing derived networks. Thus, we expect that there are more fundamental differences between contact and shear modulus fluctuations in disordered solids, that warrant future study.

\subsection{\label{sec:chi_z_chi_mu}Direct comparison of $\chi_z$ and $\chi_\mu$ in harmonic sphere packings}

\begin{figure}[h]
\includegraphics{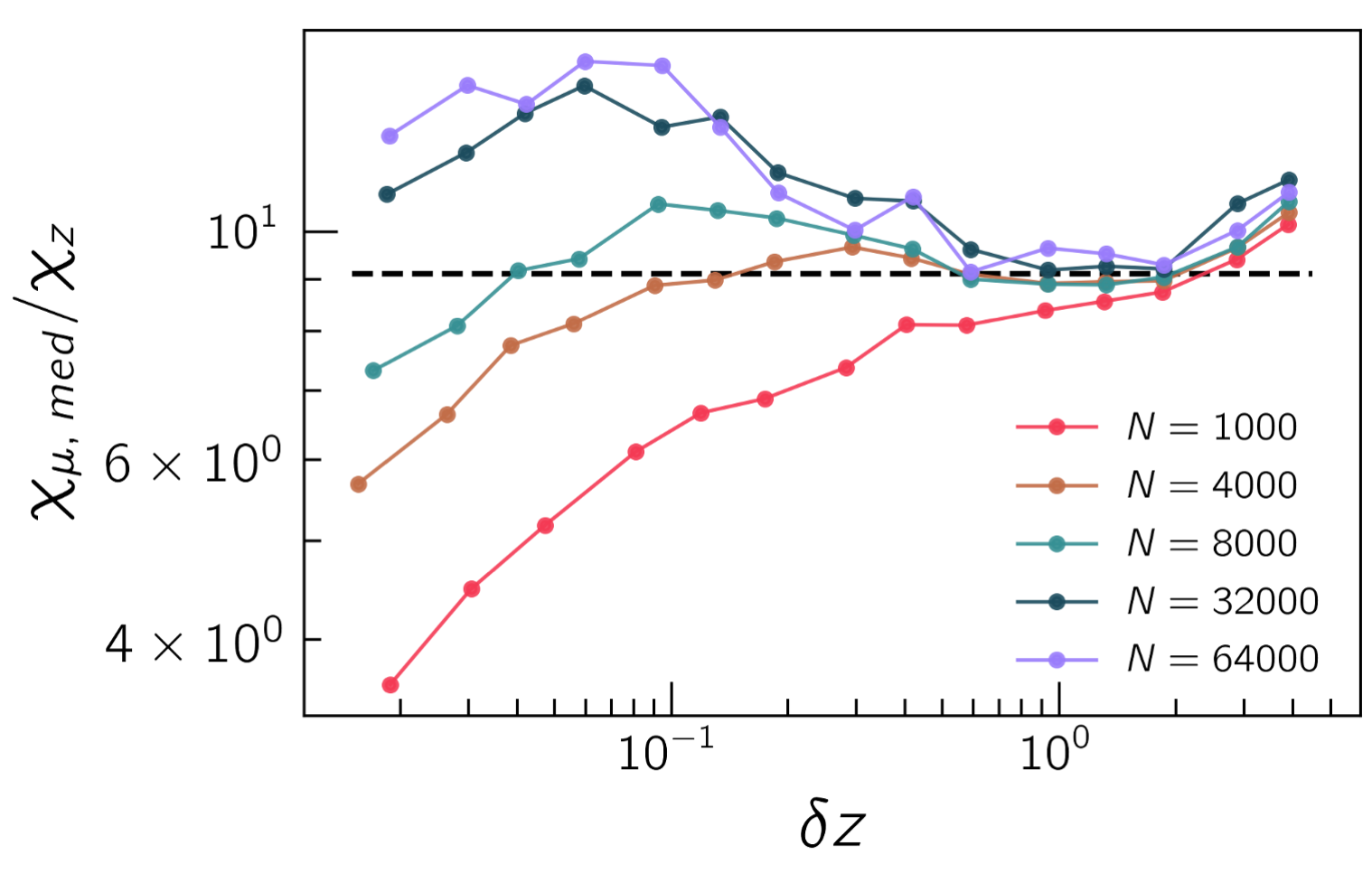}
\caption{\label{fig:chig_chiz} Ratio $\chi_{\mu, \text{med}}/\chi_z$ for harmonic sphere packings at all system sizes ($N$ increasing from bottom to top) as a function of $\delta z$. The dashed horizontal line at $\chi_{\mu, \text{med}}/\chi_z \sim 9$ is a guide to the eye, depicting the realationship $\chi_{\mu} \sim \chi_{z}$.}
\end{figure}

Lastly, we test a main hypothesis of this work, that $\chi_z \sim \chi_{\mu}$. Fig.~\ref{fig:chig_chiz} shows the ratio $\chi_{\mu, \text{med}}/\chi_z$ as a function of $\delta z$ for our ensembles of harmonic sphere packings. If $\chi_z \sim \chi_{\mu}$ holds in general, this ratio should be constant for large ranges in $\delta z$ and $N$. However, $\chi_{\mu, \text{med}}/\chi_z$ deviates significantly from constant for all system sizes. Even for the largest system sizes $N = 32000$ and $N=64000$, there is a systematic dependence of $\chi_{\mu, \text{med}}/\chi_z$ on $\delta z$. Particularly, the ratio between the two disorder quantifiers grows in the finite size regime, decreases in the unjamming regime, and is relatively constant in the glassy regime. We emphasize here that the behavior of the ratio $\chi_{\mu, \text{med}}/\chi_z$ in the unjamming regime ($N \gtrsim 8000$ and $10^{-1} \lesssim \delta z \lesssim 1$) is consistent with the interpretation that $\chi_z(\delta z)$ and $\chi_\mu(\delta z)$ display different scaling exponents $\gamma$ there. These observations are significant, as they provide intuition as to which principles connecting $\chi_{\mu}$ to other mechanical and vibrational properties (see Sec.~\ref{sec:disorder_quant}) can be extended to $\chi_z$. 

\section{\label{sec:discussion} Discussion and outlook}

In this work, we investigated the scaling properties of sample-to-sample contact and elastic modulus fluctuations in disordered solids (harmonic sphere packings and diluted spring networks) approaching the unjamming transition. Our numerical results support three conclusions that have important implications for understanding critical behavior near the transition as well as related vibrational and mechanical features of low-coordination solids. First and foremost, we identified three scaling regimes and two crossovers in the disorder quantifiers $\chi_z(N, \delta z)$ and $\chi_\mu(N, \delta z)$: the finite-size regime at low-$\delta z$ and small $N$, the unjamming regime at low-intermediate $\delta z$ and intermediate-large $N$, and the system size independent glassy regime at large $\delta z$. Even though they likely have significant effects on measurements of the diverging coordination length scale $\xi$ and corresponding mechanical behavior, these regimes in $N$ and $\delta z$ have not been thoroughly accounted for in previous literature. Second, in our investigation of the scaling of $\chi_z$ with $\delta z$ in the unjamming regime, we determined that the value of the scaling exponent $\gamma \approx -2/5$ differs considerably from $\gamma = -1/2$, the value predicted by EMT, observed widely in existing studies, and observed in this study (Sec.~\ref{sec:springs} above) for diluted spring networks \cite{wyart_scaling_2010, silbert_vibrations_2005, lerner_quasilocalized_2018}. While more research is required to understand this discrepancy, we propose that the difference in scaling behavior is related to our third major finding, that there are fundamental differences in the behavior of contact and shear modulus fluctuations in sphere packings under decompression. In other words, even though the means of the sample-to-sample distributions of $\mu$ and $\delta z$ scale together near unjamming, the variances do not. 

Relevant to our findings here, Ref.~\onlinecite{goodrich_jamming_2014} briefly examines contact and shear modulus fluctuations as a function of system size and proximity to unjamming in their discussion of the effects of finite size systems on elastic response. Particularly, the authors investigate distributions of $\delta z$ and $\mu$ in ensembles of harmonic packings by computing $\sigma_Z/ \langle \delta z \rangle $ and $\sigma_\mu / \langle \mu \rangle $ as a function of $pN^2$. They conclude that the scaling of both quantities is consistent with $\sigma_X / \langle X \rangle \sim (pN^2)^{-1/4}$ (or equivalently $\chi_X \sim \delta z^{-1/2}$) for some range in $pN^2$. However, we note that, upon close examination, there are subtle differences in the behavior of $\sigma_Z/ \langle \delta z \rangle $ and $\sigma_\mu / \langle \mu \rangle $  that seem to support our conclusions. First, the scaling of $\sigma_Z/ \langle \delta z \rangle $  with $pN^2$ deviates slightly from $(pN^2)^{-1/4}$, while $\sigma_\mu / \langle \mu \rangle $  seems better characterized by that power law. Further, the characteristics of the two standard deviations differ significantly at low $pN^2$, which is consistent with our observation that the contact and shear modulus fluctuations exhibit different finite size effects at low coordination, and that $\chi_z \sim \chi_\mu$ does \textit{not} hold in general. 

Our results presented above motivate several directions for future research. For example, subsequent studies might investigate interpretations and consequences of the unique scaling behavior of $\chi_z$ in the unjamming regime. More specifically, the relation $\xi \sim \chi_{\mu} \sim \delta z^{-1/2}$ implies that there may exist a previously unstudied unjamming length scale (which for the sake of discussion we refer to as $\xi_z$) that scales as $\xi_z \sim \chi_z \sim \delta z^{-2/5}$. If such a length scale associated with contact fluctuations did exist, future work should focus on its physical meaning. If $\xi \sim \delta z^{-1/2}$ corresponds to the size of glassy defects and is reflected in the scaling of the disorder quantifier $\chi_{\mu}(\delta z)$, what does $\xi_z \sim \delta z^{-2/5}$ correspond to that is reflected in the scaling of $\chi_z(\delta z)$? Overall, as noted in Sec.~\ref{sec:regime3}, it would be valuable for future work to further investigate the relationships between disorder quantifiers $\chi_X$ and diverging length scales approaching the unjamming transition. Alternatively, model details such as microscopic interaction potential and decompression protocol may contribute to discrepancies in the unjamming scaling behavior of $\chi_X(\delta z)$ for different observables.  

\begin{figure*}[ht]
\includegraphics[width = 4in]{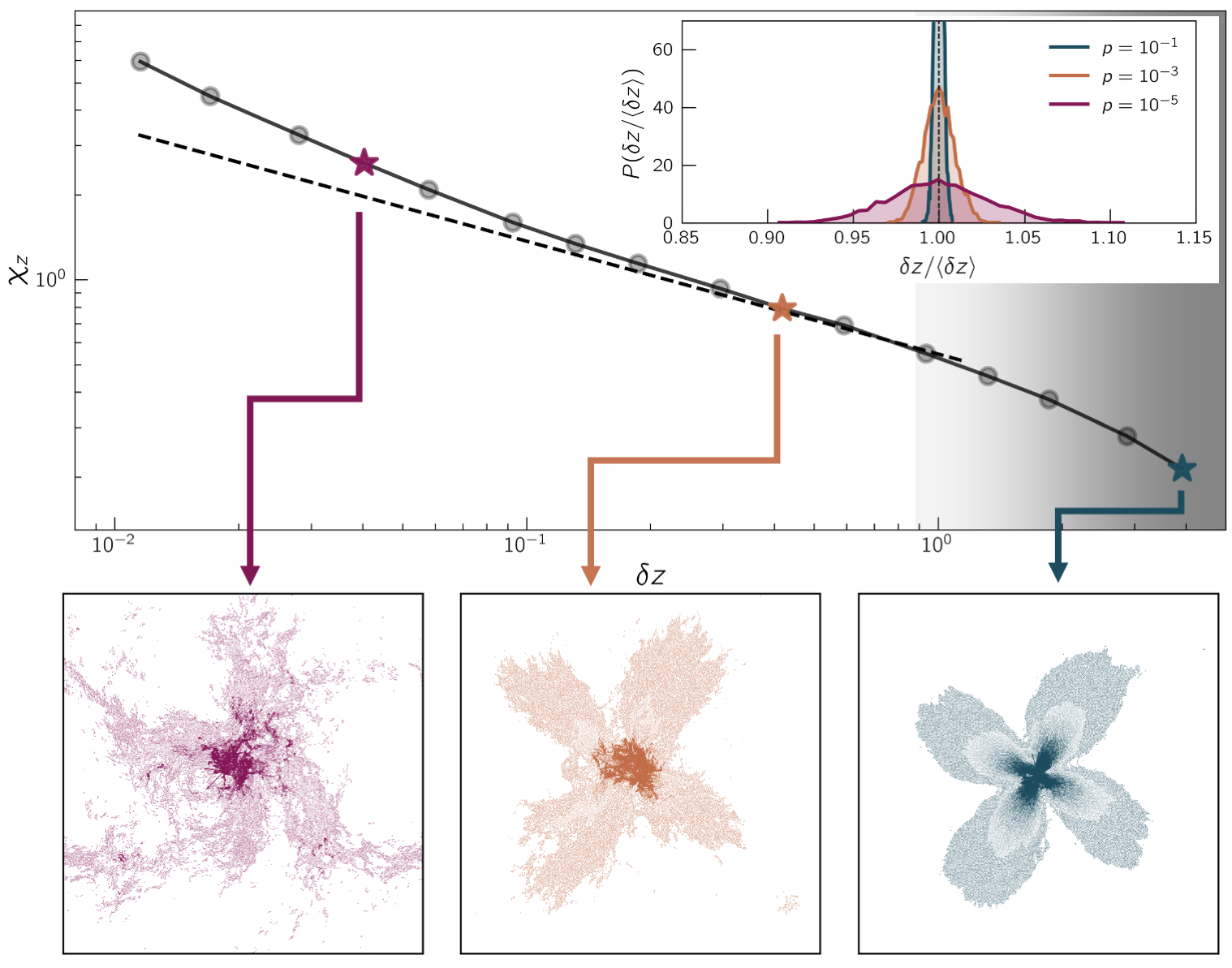}
\caption{\label{fig:highlight} Data from Figs.~\ref{fig:hists}, \ref{fig:chiz_regimes}, and \ref{fig:dipoles_all} re-plotted to summarize the main findings of this work. Each color corresponds to systems prepared at different pressures ranging from $p = 10^{-5}$ (finite-size/unjamming regime, leftmost displacement field and starred data point, broadest distribution (inset)) to $p = 10^{-1}$ (glassy regime, rightmost displacement field and starred data point, narrowest distribution (inset)).}
\end{figure*}

As we have shown, the relative behavior of $\chi_z$ and $\chi_\mu$ as a function of $N$ and $\delta z$ differs significantly from the naive expectation, $\chi_z \sim \chi_\mu$, based on the fact that the means of $\mu$ and $\delta z$ are proportional. This motivates further characterization of the differences between sample-to-sample distributions of shear moduli and coordination numbers in disordered solids approaching unjamming. Notably, Ref.~\onlinecite{kapteijns_elastic_2021} found that such shear modulus distributions exhibit heavy leftward tails (visible in Fig.~\ref{fig:hists}a above) of the form $P(\mu;N) \sim N^{-3/2} (\langle \mu \rangle - \mu)^{-7/2}$. Considering the dependence of the nonaffine contribution to the shear modulus on the vibrational density of states (see Ref.~\onlinecite{hentschel_athermal_2011}), the authors of Ref.~\onlinecite{kapteijns_elastic_2021} attribute the behavior of these tails to the existence of low-frequency non-phononic (quasilocalized) excitations in disordered solids. Conversely, there is not as direct of a connection between the vibrational spectra of glasses and their contact statistics, and heavy leftward tails are not visible in Fig.~\ref{fig:hists}b above. Future work investigating connections between $\chi_z$ and $\chi_\mu$ as quantifiers of mechanical disorder would be useful, as $\chi_\mu$ is directly related to other quantities of interest that can be measured in sound attenuation experiments.

\begin{acknowledgments}
J.A.G.~thanks Sadjad Arzash, R. Cameron Dennis, and Tyler A. Hain for helpful conversations, and David Richard for significant contributions to the analysis code used in this work. J.A.G.~acknowledges use of Syracuse University's computational resources (OrangeGrid) and related support from Syracuse's research computing team.  E.L.~acknowledges Support from the NWO (Vidi grant no.~680-47-554/3259). M.L.M.~and J.G.~were supported by NSF DMR 1951921 and Simons Foundation \#454947.
\end{acknowledgments}

\section*{Author Contributions Statement}
All authors contributed to the overall conceptualization of this work. J.A.G.~and E.L.~wrote and developed the simulation and analysis code. J.A.G.~and E.L.~ran the simulation and analysis code. All authors contributed to the interpretation of the results. J.A.G.~prepared the figures and original draft of the manuscript. All authors contributed to reviewing and editing the manuscript. 

\appendix

\section{Summary Figure}
Fig. \ref{fig:highlight} summarizes the results of much of the analysis and discussion presented above. In modeled disordered solids approaching the unjamming transition (vanishing $p$ and $\delta z$), the sample-to-sample distribution of excess contact number $\delta z$ broadens significantly and $\chi_Z$ quickly increases. Simultaneously, the core size $\xi$ of low-energy excitations grows with decreasing pressure as the corresponding displacement fields (represented by dipolar response fields) become less characteristic of local elasticity.

\section{\label{app:chiz_reduced}Additional collapses of $\chi_z$ in the unjamming regime}

\begin{figure}[h]
\includegraphics{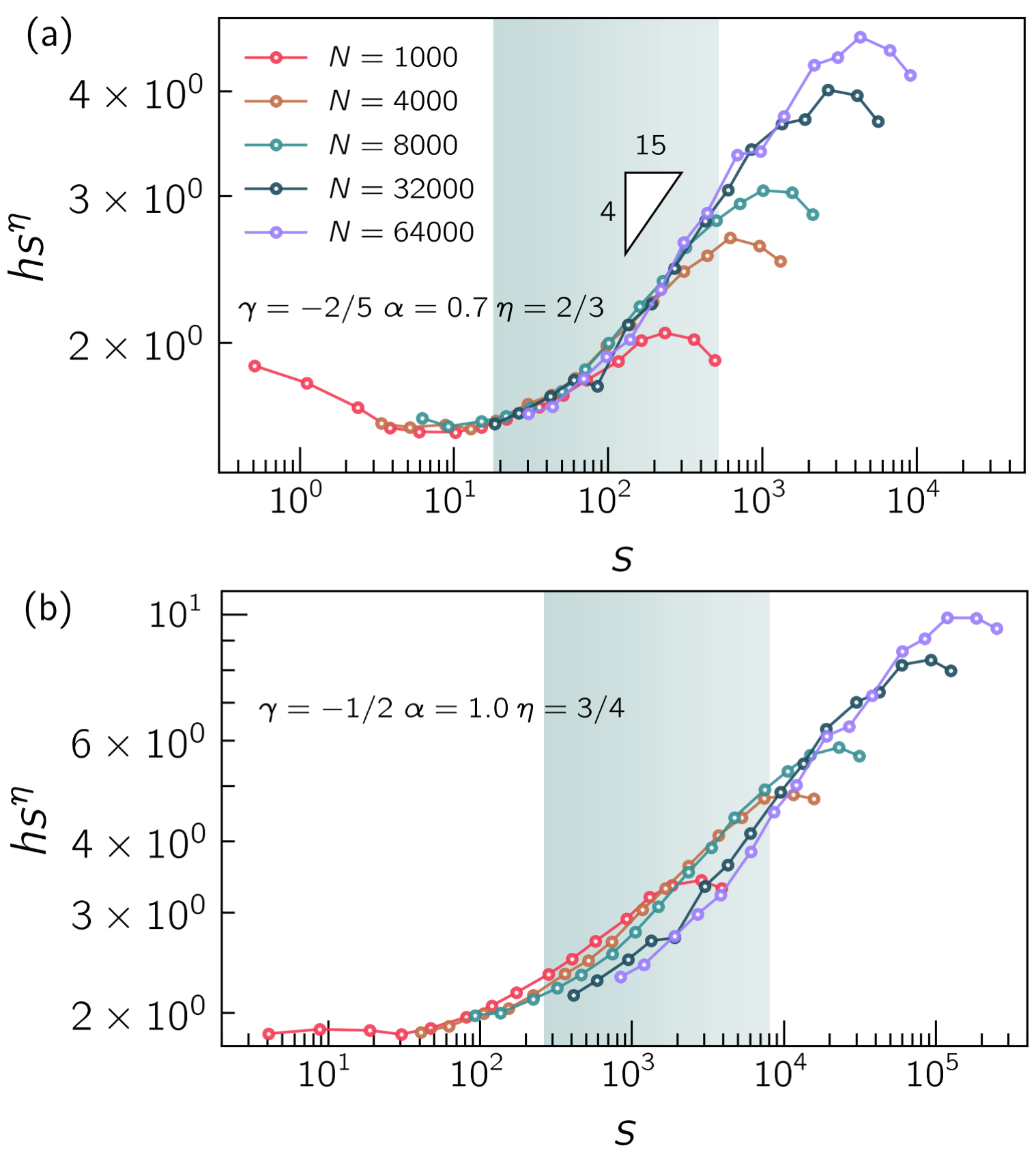}
\caption{\label{fig:chiz_reduced} Reduced representations of $\chi_z(N, \delta z)$ further comparing values of $\gamma$, similarly to Fig.~\ref{fig:chiz_collapse} in the main text. (a) Scaling collapse according to the exponent $\gamma \approx -2/5$ and the parameters $\alpha \approx 0.7$ and $\eta \approx 2/3$. The scaling of the data in the unjamming regime is consistent with $\gamma + \eta \approx 4/15$, as shown by the scale bar. (b) Scaling collapse according to the exponent $\gamma \approx -1/2$ and the scaling parameters $\alpha \approx 1.0$ and $\eta \approx 3/4$. The data does not collapse as well in the unjamming regime as the rescaling in panel (a). The shaded region depicts the approximate extent of the unjamming regime for the system size $N=8000$, where we expect the data collapse and $s^{\gamma + \eta}$ scaling to be well-represented.}
\end{figure}

To provide additional support for our measurement of the exponent $\vert \gamma \vert < 1/2$, Fig.~\ref{fig:chiz_reduced} shows a alternative representation of the same $\chi_z(N,\delta z)$ data as Fig.~\ref{fig:chiz_collapse} in the main text, in terms of the quantities $h = N^{\alpha \gamma} \chi_z$ and $s = N^{\alpha} \delta z$. Observing that both panels of  Fig.~\ref{fig:chiz_collapse} exhibit the scaling $h \sim s^{-\eta}$ in the finite-size regime, we plot $hs^{\eta}$ vs. $s$ in Fig.~\ref{fig:chiz_reduced}. With this rescaling, we expect the data in the finite-size regime to stay relatively constant as a function of $s$, and the data in the unjamming regime to scale as $s^{\gamma + \eta}$. Viewing the quality of the scaling collapses in panels (a) and (b) of Fig.~\ref{fig:chiz_reduced}, we again conclude that $\chi_z$ scales as $\chi_z \sim \delta z^{\gamma}$ with $\gamma \approx -2/5$ in the unjamming regime. The unjamming regime for the system size $N=8000$ is highlighted in both panels of Fig.~\ref{fig:chiz_reduced}. We emphasize again that the collapses due to this rescaling and that in Fig.~\ref{fig:chiz_collapse} are only expected to work up to the onset of the ($N$-independent) glassy regime for each system size.

\section{\label{app:chi_mu_packings}$\chi_\mu$ for harmonic packings}

\begin{figure}[h]
\includegraphics{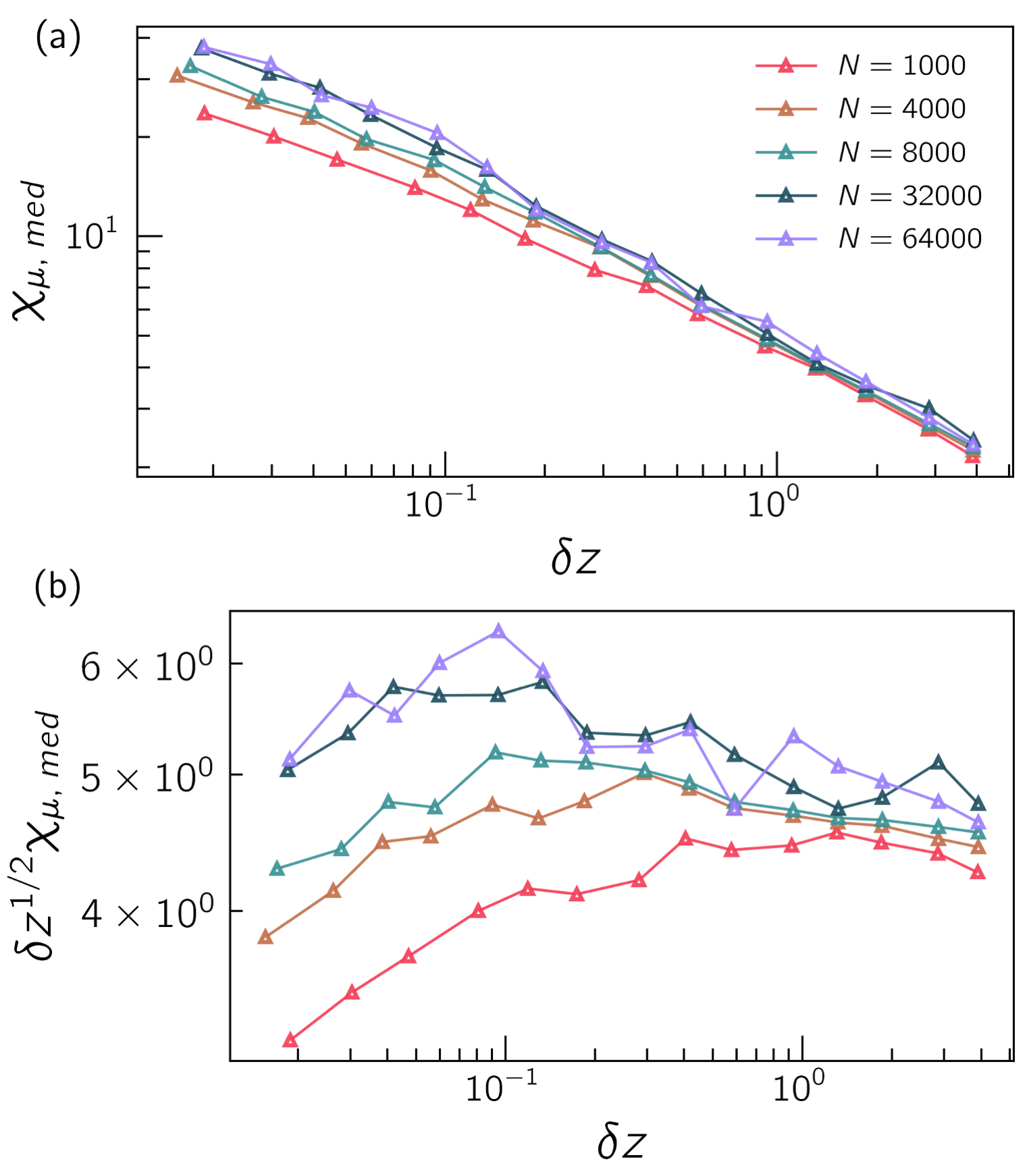}
\caption{\label{fig:chimu_raw} The shear modulus disorder quantifier $\chi_{\mu, \text{med}}$ computed using the median method and measured in our ensembles of harmonic sphere packings approaching the unjamming transition. (a) The raw $\chi_{\mu, \text{med}}(\delta z)$ data plotted for all system sizes (increasing $N$ from bottom to top). (b) The same data as panel (a) rescaled by $\delta z^{-1/2}$ on the vertical axis to check the correspondence to the scaling relationship $\chi_{\mu} \sim \delta z^{-1/2}$.}
\end{figure}

As mentioned in the main text, due to large shear modulus fluctuations in our ensembles of packings approaching the unjamming transition, the functional behavior of $\chi_{\mu}(N, \delta z)$ is too noisy to determine robust (power law) scaling relationships (distinctly to our results presented above for $\chi_z$ in packings and $\chi_\mu$ in spring networks). Fig.~\ref{fig:chimu_raw}a shows $\chi_{\mu, \text{med}}(N, \delta z)$ for the same ensembles of packings as Fig.~\ref{fig:chiz_regimes}. The scaling regimes discussed in the main text (finite size, unjamming, and glassy) are likely still present, but it is difficult to discern from this data set. Still, to check the general scaling behavior of $\chi_{\mu, \text{med}}$ in the unjamming regime, we plot $\chi_{\mu, \text{med}}/\delta z^{-1/2}$ as a function of $\delta z$ in Fig.~\ref{fig:chimu_raw}b. Again, the behavior is quite noisy, but overall consistent with the scaling relationship $\chi_\mu \sim \delta z^{-1/2}$.

\section{\label{app:dipole_responses} Dipole response fields in $\dbar = 2$ harmonic packings}

\begin{figure}[h]
\includegraphics{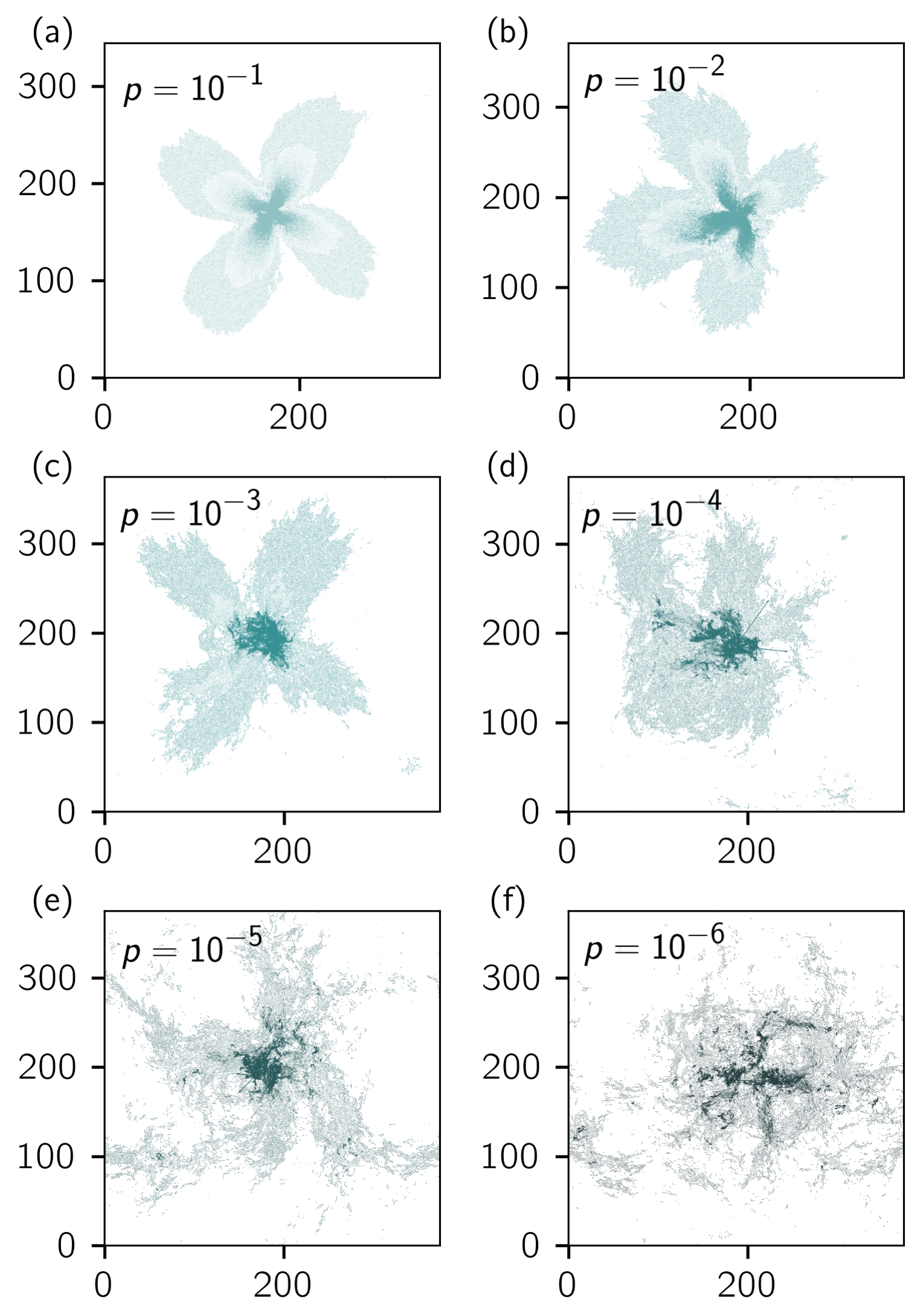}
\caption{\label{fig:dipoles_all} Displacement fields that result from applied force dipoles in harmonic packings with $\dbar = 2$ and $N = 102400$, similarly to Fig.~\ref{fig:dipole_res} in the main text. Panels (a)-(f) show these displacement fields for packings prepared at the specified overall pressures $p \in \left[10^{-6}, 10^{-1} \right]$. Note that the fields become significantly less localized and less characterizable as quadrupolar elastic fields with decreasing pressure.}
\end{figure}

To more specifically depict the growth of the length scale $\xi$ approaching the unjamming transition, Fig.\ref{fig:dipoles_all} shows example displacement response fields in two-dimensional harmonic packings subject to applied force dipoles. As described in Sec.~\ref{sec:regime3}, these displacement fields are generally characterized by disordered cores of size $\xi$ decorated by elastic decaying fields. This structure is the most discernible deep in the glassy regime, where continuum elasticity is the most relevant to the macroscopic behavior. As the overall pressure $p \in \left[10^{-6}, 10^{-1} \right]$ of the packings decreases, the core size $\xi$ drastically increases and the displacement fields exhibit the disordered, anomalous behavior usually associated with the unjamming transition. 

\section{\label{app:model_sys}Comparison of model systems}

In this Appendix, we directly compare the scaling properties (at constant system size $N$ and varying excess coordination $\delta z$) of all of the disorder quantifiers discussed thus far: $\chi_{z}$ in packings, $\chi_{\mu}$ in packings, and $\chi_{\mu}$ in diluted spring networks. Additionally, we examine $\chi_{\mu}$ in packing-derived spring networks to test the hypothesis proposed in Sec.~\ref{sec:springs} above, that the observed differences in scaling behavior between $\chi_{z}(N, \delta z)$ in packings and $\chi_{\mu}(N, \delta z)$ in diluted spring networks can be explained by the presence or absence of considerable structural rearrangements under decompression. To begin, we briefly describe our method for producing packing-derived networks. 

\subsection{Preparation of packing-derived networks}

In comparison to the harmonic sphere packings and diluted spring networks described above, we study packing-derived networks. Similarly to the protocol for initializing diluted spring networks, the packing-derived networks are comprised of unstressed Hookean springs and adopt the structure of our configurations of harmonic spheres at varying overall pressure $p$. Thus, as particulate rearrangements occur in the modeled glasses under decompression, the same changes in coordination are present in the corresponding spring networks. In our scaling analysis of shear modulus fluctuations approaching unjamming, we examine the ensemble of networks that corresponds to the packings with $N = 8000$ and $N_{\text{ens}} = 5000$.

\subsection{Disorder quantifiers in spring networks and harmonic packings}

\begin{figure}[h]
\includegraphics{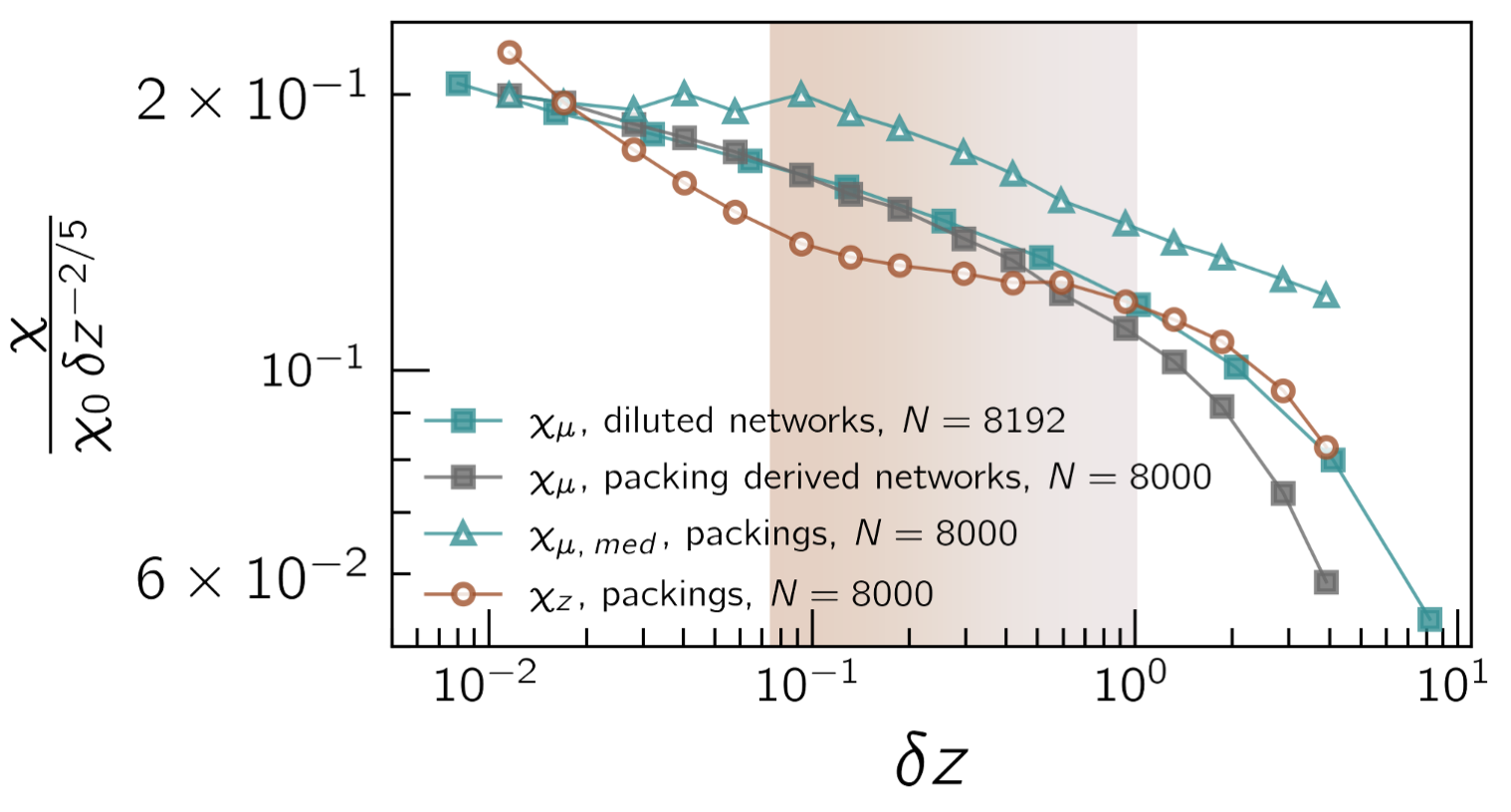}
\caption{\label{fig:models_scaling} Shear modulus disorder quantifier $\chi_\mu$ as a function of $\delta z$ with $N\sim8000$ for ensembles of sphere packings, packing-derived networks, and diluted networks. Contact number disorder quantifier $\chi_z$ for the same ensemble of sphere packings. The data for each curve are rescaled by the value $\chi_0 \sim \chi(p \sim 2 \times 10^{-6})$ for visualization purposes. Additionally, the y-axis is rescaled by $\delta z^{\gamma \approx -2/5}$. The shaded region highlights the approximate extent of the unjamming regime for $\chi_z$ measured in packings. Notably, the rescaled $\chi_z$ data remains relatively constant in this regime, contrasting all $\chi_\mu$ measurements in the same range of $\delta z$.}
\end{figure}

Fig.~\ref{fig:models_scaling} shows the shear modulus disorder quantifier $\chi_\mu(\delta z)$ for our ensembles of harmonic packings, diluted spring networks, and packing-derived spring networks with $N \approx 8000$. Additionally, it shows the contact number quantifier $\chi_z(\delta z)$ for the ensemble of harmonic packings with $N = 8000$. Since each $\chi(\delta z)$ curve has a different overall scale, we rescale them by their value at $p \sim 2 \times 10^{-6}$ (denoted $\chi_0$) for ease of comparison. Last, we rescale the vertical axis by $\delta z ^{-2/5}$, which is the approximate scaling of $\chi_z$ in the unjamming regime as described in the main text. With these rescalings, the three scaling regimes in $\chi_z(\delta z)$ are clear to see. More specifically, the rescaled $\chi_z$ curve is horizontal in the shaded region of the figure, $8\times10^{-2} \lesssim \delta z \lesssim 1$, while all of the rescaled $\chi_\mu$ curves have a noticeable slope. This provides further evidence for our claims that i) $\chi_z \sim \delta z^{\gamma}$ with $\vert \gamma \vert < 1/2$ in the unjamming regime and ii) $\chi_z \sim \chi_{\mu}$ does not hold in general. 

Now comparing the behavior of $\chi_{\mu}(N, \delta z)$ for the two different spring network preparations, we notice that the scaling behavior is very similar in the finite size and unjamming regimes. The most noticeable difference between $\chi_\mu$'s behavior for diluted vs. packing-derived networks occurs in the crossover to the glassy regime, which begins slightly sooner in $\delta z$ for the packing-derived networks. Overall, we conclude that the coordination changes that occur in the packing-derived networks do not contribute significantly to the scaling behavior (i.e. the measurement of the exponent $\gamma$) in the unjamming regime.

\newpage 

\bibliography{apssamp}

\begin{thebibliography}{54}%
\makeatletter
\providecommand \@ifxundefined [1]{%
 \@ifx{#1\undefined}
}%
\providecommand \@ifnum [1]{%
 \ifnum #1\expandafter \@firstoftwo
 \else \expandafter \@secondoftwo
 \fi
}%
\providecommand \@ifx [1]{%
 \ifx #1\expandafter \@firstoftwo
 \else \expandafter \@secondoftwo
 \fi
}%
\providecommand \natexlab [1]{#1}%
\providecommand \enquote  [1]{``#1''}%
\providecommand \bibnamefont  [1]{#1}%
\providecommand \bibfnamefont [1]{#1}%
\providecommand \citenamefont [1]{#1}%
\providecommand \href@noop [0]{\@secondoftwo}%
\providecommand \href [0]{\begingroup \@sanitize@url \@href}%
\providecommand \@href[1]{\@@startlink{#1}\@@href}%
\providecommand \@@href[1]{\endgroup#1\@@endlink}%
\providecommand \@sanitize@url [0]{\catcode `\\12\catcode `\$12\catcode `\&12\catcode `\#12\catcode `\^12\catcode `\_12\catcode `\%12\relax}%
\providecommand \@@startlink[1]{}%
\providecommand \@@endlink[0]{}%
\providecommand \url  [0]{\begingroup\@sanitize@url \@url }%
\providecommand \@url [1]{\endgroup\@href {#1}{\urlprefix }}%
\providecommand \urlprefix  [0]{URL }%
\providecommand \Eprint [0]{\href }%
\providecommand \doibase [0]{https://doi.org/}%
\providecommand \selectlanguage [0]{\@gobble}%
\providecommand \bibinfo  [0]{\@secondoftwo}%
\providecommand \bibfield  [0]{\@secondoftwo}%
\providecommand \translation [1]{[#1]}%
\providecommand \BibitemOpen [0]{}%
\providecommand \bibitemStop [0]{}%
\providecommand \bibitemNoStop [0]{.\EOS\space}%
\providecommand \EOS [0]{\spacefactor3000\relax}%
\providecommand \BibitemShut  [1]{\csname bibitem#1\endcsname}%
\let\auto@bib@innerbib\@empty
\bibitem [{\citenamefont {Behringer}\ and\ \citenamefont {Chakraborty}(2018)}]{behringer_physics_2018}%
  \BibitemOpen
  \bibfield  {author} {\bibinfo {author} {\bibfnamefont {R.~P.}\ \bibnamefont {Behringer}}\ and\ \bibinfo {author} {\bibfnamefont {B.}~\bibnamefont {Chakraborty}},\ }\bibfield  {title} {\bibinfo {title} {The physics of jamming for granular materials: a review},\ }\href {https://doi.org/10.1088/1361-6633/aadc3c} {\bibfield  {journal} {\bibinfo  {journal} {Reports on Progress in Physics}\ }\textbf {\bibinfo {volume} {82}},\ \bibinfo {pages} {012601} (\bibinfo {year} {2018})},\ \bibinfo {note} {publisher: IOP Publishing}\BibitemShut {NoStop}%
\bibitem [{\citenamefont {Liu}\ and\ \citenamefont {Nagel}(1998)}]{liu_jamming_1998}%
  \BibitemOpen
  \bibfield  {author} {\bibinfo {author} {\bibfnamefont {A.~J.}\ \bibnamefont {Liu}}\ and\ \bibinfo {author} {\bibfnamefont {S.~R.}\ \bibnamefont {Nagel}},\ }\bibfield  {title} {\bibinfo {title} {Jamming is not just cool any more},\ }\href {https://doi.org/http://dx.doi.org.libezproxy2.syr.edu/10.1038/23819} {\bibfield  {journal} {\bibinfo  {journal} {Nature; London}\ }\textbf {\bibinfo {volume} {396}},\ \bibinfo {pages} {21} (\bibinfo {year} {1998})},\ \bibinfo {note} {num Pages: 21-22 Place: London, United States, London Publisher: Nature Publishing Group}\BibitemShut {NoStop}%
\bibitem [{\citenamefont {Péter}\ \emph {et~al.}(2018)\citenamefont {Péter}, \citenamefont {Libál}, \citenamefont {Reichhardt},\ and\ \citenamefont {Reichhardt}}]{peter_crossover_2018}%
  \BibitemOpen
  \bibfield  {author} {\bibinfo {author} {\bibfnamefont {H.}~\bibnamefont {Péter}}, \bibinfo {author} {\bibfnamefont {A.}~\bibnamefont {Libál}}, \bibinfo {author} {\bibfnamefont {C.}~\bibnamefont {Reichhardt}},\ and\ \bibinfo {author} {\bibfnamefont {C.~J.~O.}\ \bibnamefont {Reichhardt}},\ }\bibfield  {title} {\bibinfo {title} {Crossover from {Jamming} to {Clogging} {Behaviours} in {Heterogeneous} {Environments}},\ }\href {https://doi.org/10.1038/s41598-018-28256-6} {\bibfield  {journal} {\bibinfo  {journal} {Scientific Reports}\ }\textbf {\bibinfo {volume} {8}},\ \bibinfo {pages} {10252} (\bibinfo {year} {2018})},\ \bibinfo {note} {number: 1 Publisher: Nature Publishing Group}\BibitemShut {NoStop}%
\bibitem [{\citenamefont {Nagatani}(2002)}]{nagatani_physics_2002}%
  \BibitemOpen
  \bibfield  {author} {\bibinfo {author} {\bibfnamefont {T.}~\bibnamefont {Nagatani}},\ }\bibfield  {title} {\bibinfo {title} {The physics of traffic jams},\ }\href {https://doi.org/10.1088/0034-4885/65/9/203} {\bibfield  {journal} {\bibinfo  {journal} {Reports on Progress in Physics}\ }\textbf {\bibinfo {volume} {65}},\ \bibinfo {pages} {1331} (\bibinfo {year} {2002})}\BibitemShut {NoStop}%
\bibitem [{\citenamefont {Lawson-Keister}\ and\ \citenamefont {Manning}(2021)}]{lawson-keister_jamming_2021}%
  \BibitemOpen
  \bibfield  {author} {\bibinfo {author} {\bibfnamefont {E.}~\bibnamefont {Lawson-Keister}}\ and\ \bibinfo {author} {\bibfnamefont {M.~L.}\ \bibnamefont {Manning}},\ }\bibfield  {title} {\bibinfo {title} {Jamming and arrest of cell motion in biological tissues},\ }\href {https://doi.org/10.1016/j.ceb.2021.07.011} {\bibfield  {journal} {\bibinfo  {journal} {Current Opinion in Cell Biology}\ }\bibinfo {series} {Cell {Dynamics}},\ \textbf {\bibinfo {volume} {72}},\ \bibinfo {pages} {146} (\bibinfo {year} {2021})}\BibitemShut {NoStop}%
\bibitem [{\citenamefont {Garcimartín}\ \emph {et~al.}(2015)\citenamefont {Garcimartín}, \citenamefont {Pastor}, \citenamefont {Ferrer}, \citenamefont {Ramos}, \citenamefont {Martín-Gómez},\ and\ \citenamefont {Zuriguel}}]{garcimartin_flow_2015}%
  \BibitemOpen
  \bibfield  {author} {\bibinfo {author} {\bibfnamefont {A.}~\bibnamefont {Garcimartín}}, \bibinfo {author} {\bibfnamefont {J.~M.}\ \bibnamefont {Pastor}}, \bibinfo {author} {\bibfnamefont {L.~M.}\ \bibnamefont {Ferrer}}, \bibinfo {author} {\bibfnamefont {J.~J.}\ \bibnamefont {Ramos}}, \bibinfo {author} {\bibfnamefont {C.}~\bibnamefont {Martín-Gómez}},\ and\ \bibinfo {author} {\bibfnamefont {I.}~\bibnamefont {Zuriguel}},\ }\bibfield  {title} {\bibinfo {title} {Flow and clogging of a sheep herd passing through a bottleneck},\ }\href {https://doi.org/10.1103/PhysRevE.91.022808} {\bibfield  {journal} {\bibinfo  {journal} {Physical Review E}\ }\textbf {\bibinfo {volume} {91}},\ \bibinfo {pages} {022808} (\bibinfo {year} {2015})}\BibitemShut {NoStop}%
\bibitem [{\citenamefont {Liu}\ and\ \citenamefont {Nagel}(2010)}]{liu_jamming_2010}%
  \BibitemOpen
  \bibfield  {author} {\bibinfo {author} {\bibfnamefont {A.~J.}\ \bibnamefont {Liu}}\ and\ \bibinfo {author} {\bibfnamefont {S.~R.}\ \bibnamefont {Nagel}},\ }\bibfield  {title} {\bibinfo {title} {The {Jamming} {Transition} and the {Marginally} {Jammed} {Solid}},\ }\href {https://doi.org/10.1146/annurev-conmatphys-070909-104045} {\bibfield  {journal} {\bibinfo  {journal} {Annual Review of Condensed Matter Physics}\ }\textbf {\bibinfo {volume} {1}},\ \bibinfo {pages} {347} (\bibinfo {year} {2010})}\BibitemShut {NoStop}%
\bibitem [{\citenamefont {Hecke}(2009)}]{hecke_jamming_2009}%
  \BibitemOpen
  \bibfield  {author} {\bibinfo {author} {\bibfnamefont {M.~v.}\ \bibnamefont {Hecke}},\ }\bibfield  {title} {\bibinfo {title} {Jamming of soft particles: geometry, mechanics, scaling and isostaticity},\ }\href {https://doi.org/10.1088/0953-8984/22/3/033101} {\bibfield  {journal} {\bibinfo  {journal} {Journal of Physics: Condensed Matter}\ }\textbf {\bibinfo {volume} {22}},\ \bibinfo {pages} {033101} (\bibinfo {year} {2009})}\BibitemShut {NoStop}%
\bibitem [{\citenamefont {Parisi}\ \emph {et~al.}(2020)\citenamefont {Parisi}, \citenamefont {Urbani},\ and\ \citenamefont {Zamponi}}]{parisi2020theory}%
  \BibitemOpen
  \bibfield  {author} {\bibinfo {author} {\bibfnamefont {G.}~\bibnamefont {Parisi}}, \bibinfo {author} {\bibfnamefont {P.}~\bibnamefont {Urbani}},\ and\ \bibinfo {author} {\bibfnamefont {F.}~\bibnamefont {Zamponi}},\ }\href@noop {} {\emph {\bibinfo {title} {Theory of simple glasses: exact solutions in infinite dimensions}}}\ (\bibinfo  {publisher} {Cambridge University Press},\ \bibinfo {year} {2020})\BibitemShut {NoStop}%
\bibitem [{\citenamefont {M{\"u}ller}\ and\ \citenamefont {Wyart}(2015)}]{muller2015marginal}%
  \BibitemOpen
  \bibfield  {author} {\bibinfo {author} {\bibfnamefont {M.}~\bibnamefont {M{\"u}ller}}\ and\ \bibinfo {author} {\bibfnamefont {M.}~\bibnamefont {Wyart}},\ }\bibfield  {title} {\bibinfo {title} {Marginal stability in structural, spin, and electron glasses},\ }\href@noop {} {\bibfield  {journal} {\bibinfo  {journal} {Annu. Rev. Condens. Matter Phys.}\ }\textbf {\bibinfo {volume} {6}},\ \bibinfo {pages} {177} (\bibinfo {year} {2015})}\BibitemShut {NoStop}%
\bibitem [{\citenamefont {Charbonneau}\ \emph {et~al.}(2015)\citenamefont {Charbonneau}, \citenamefont {Corwin}, \citenamefont {Parisi},\ and\ \citenamefont {Zamponi}}]{charbonneau_jamming_2015}%
  \BibitemOpen
  \bibfield  {author} {\bibinfo {author} {\bibfnamefont {P.}~\bibnamefont {Charbonneau}}, \bibinfo {author} {\bibfnamefont {E.~I.}\ \bibnamefont {Corwin}}, \bibinfo {author} {\bibfnamefont {G.}~\bibnamefont {Parisi}},\ and\ \bibinfo {author} {\bibfnamefont {F.}~\bibnamefont {Zamponi}},\ }\bibfield  {title} {\bibinfo {title} {Jamming {Criticality} {Revealed} by {Removing} {Localized} {Buckling} {Excitations}},\ }\href {https://doi.org/10.1103/PhysRevLett.114.125504} {\bibfield  {journal} {\bibinfo  {journal} {Physical Review Letters}\ }\textbf {\bibinfo {volume} {114}},\ \bibinfo {pages} {125504} (\bibinfo {year} {2015})},\ \bibinfo {note} {arXiv:1411.3975 [cond-mat]}\BibitemShut {NoStop}%
\bibitem [{\citenamefont {Franz}\ \emph {et~al.}(2015)\citenamefont {Franz}, \citenamefont {Parisi}, \citenamefont {Urbani},\ and\ \citenamefont {Zamponi}}]{franz2015universal}%
  \BibitemOpen
  \bibfield  {author} {\bibinfo {author} {\bibfnamefont {S.}~\bibnamefont {Franz}}, \bibinfo {author} {\bibfnamefont {G.}~\bibnamefont {Parisi}}, \bibinfo {author} {\bibfnamefont {P.}~\bibnamefont {Urbani}},\ and\ \bibinfo {author} {\bibfnamefont {F.}~\bibnamefont {Zamponi}},\ }\bibfield  {title} {\bibinfo {title} {Universal spectrum of normal modes in low-temperature glasses},\ }\href@noop {} {\bibfield  {journal} {\bibinfo  {journal} {Proceedings of the National Academy of Sciences}\ }\textbf {\bibinfo {volume} {112}},\ \bibinfo {pages} {14539} (\bibinfo {year} {2015})}\BibitemShut {NoStop}%
\bibitem [{\citenamefont {Lerner}(2018)}]{lerner_quasilocalized_2018}%
  \BibitemOpen
  \bibfield  {author} {\bibinfo {author} {\bibfnamefont {E.}~\bibnamefont {Lerner}},\ }\bibfield  {title} {\bibinfo {title} {Quasilocalized states of self stress in packing-derived networks},\ }\href {https://doi.org/10.1140/epje/i2018-11705-9} {\bibfield  {journal} {\bibinfo  {journal} {The European Physical Journal E}\ }\textbf {\bibinfo {volume} {41}},\ \bibinfo {pages} {93} (\bibinfo {year} {2018})}\BibitemShut {NoStop}%
\bibitem [{\citenamefont {Karimi}\ and\ \citenamefont {Maloney}(2015)}]{karimi_elasticity_2015}%
  \BibitemOpen
  \bibfield  {author} {\bibinfo {author} {\bibfnamefont {K.}~\bibnamefont {Karimi}}\ and\ \bibinfo {author} {\bibfnamefont {C.~E.}\ \bibnamefont {Maloney}},\ }\bibfield  {title} {\bibinfo {title} {Elasticity of frictionless particles near jamming},\ }\href {https://doi.org/10.1103/PhysRevE.92.022208} {\bibfield  {journal} {\bibinfo  {journal} {Physical Review E}\ }\textbf {\bibinfo {volume} {92}},\ \bibinfo {pages} {022208} (\bibinfo {year} {2015})}\BibitemShut {NoStop}%
\bibitem [{\citenamefont {Baumgarten}\ \emph {et~al.}(2017)\citenamefont {Baumgarten}, \citenamefont {Vågberg},\ and\ \citenamefont {Tighe}}]{baumgarten_nonlocal_2017}%
  \BibitemOpen
  \bibfield  {author} {\bibinfo {author} {\bibfnamefont {K.}~\bibnamefont {Baumgarten}}, \bibinfo {author} {\bibfnamefont {D.}~\bibnamefont {Vågberg}},\ and\ \bibinfo {author} {\bibfnamefont {B.~P.}\ \bibnamefont {Tighe}},\ }\bibfield  {title} {\bibinfo {title} {Nonlocal {Elasticity} near {Jamming} in {Frictionless} {Soft} {Spheres}},\ }\href {https://doi.org/10.1103/PhysRevLett.118.098001} {\bibfield  {journal} {\bibinfo  {journal} {Physical Review Letters}\ }\textbf {\bibinfo {volume} {118}},\ \bibinfo {pages} {098001} (\bibinfo {year} {2017})}\BibitemShut {NoStop}%
\bibitem [{\citenamefont {Düring}\ \emph {et~al.}(2014)\citenamefont {Düring}, \citenamefont {Lerner},\ and\ \citenamefont {Wyart}}]{during_length_2014}%
  \BibitemOpen
  \bibfield  {author} {\bibinfo {author} {\bibfnamefont {G.}~\bibnamefont {Düring}}, \bibinfo {author} {\bibfnamefont {E.}~\bibnamefont {Lerner}},\ and\ \bibinfo {author} {\bibfnamefont {M.}~\bibnamefont {Wyart}},\ }\bibfield  {title} {\bibinfo {title} {Length scales and self-organization in dense suspension flows},\ }\href {https://doi.org/10.1103/PhysRevE.89.022305} {\bibfield  {journal} {\bibinfo  {journal} {Physical Review E}\ }\textbf {\bibinfo {volume} {89}},\ \bibinfo {pages} {022305} (\bibinfo {year} {2014})},\ \bibinfo {note} {publisher: American Physical Society}\BibitemShut {NoStop}%
\bibitem [{\citenamefont {Bouzid}\ \emph {et~al.}(2013)\citenamefont {Bouzid}, \citenamefont {Trulsson}, \citenamefont {Claudin}, \citenamefont {Clément},\ and\ \citenamefont {Andreotti}}]{bouzid_nonlocal_2013}%
  \BibitemOpen
  \bibfield  {author} {\bibinfo {author} {\bibfnamefont {M.}~\bibnamefont {Bouzid}}, \bibinfo {author} {\bibfnamefont {M.}~\bibnamefont {Trulsson}}, \bibinfo {author} {\bibfnamefont {P.}~\bibnamefont {Claudin}}, \bibinfo {author} {\bibfnamefont {E.}~\bibnamefont {Clément}},\ and\ \bibinfo {author} {\bibfnamefont {B.}~\bibnamefont {Andreotti}},\ }\bibfield  {title} {\bibinfo {title} {Nonlocal {Rheology} of {Granular} {Flows} across {Yield} {Conditions}},\ }\href {https://doi.org/10.1103/PhysRevLett.111.238301} {\bibfield  {journal} {\bibinfo  {journal} {Physical Review Letters}\ }\textbf {\bibinfo {volume} {111}},\ \bibinfo {pages} {238301} (\bibinfo {year} {2013})},\ \bibinfo {note} {publisher: American Physical Society}\BibitemShut {NoStop}%
\bibitem [{\citenamefont {Wyart}\ \emph {et~al.}(2005)\citenamefont {Wyart}, \citenamefont {Nagel},\ and\ \citenamefont {Witten}}]{wyart_geometric_2005}%
  \BibitemOpen
  \bibfield  {author} {\bibinfo {author} {\bibfnamefont {M.}~\bibnamefont {Wyart}}, \bibinfo {author} {\bibfnamefont {S.~R.}\ \bibnamefont {Nagel}},\ and\ \bibinfo {author} {\bibfnamefont {T.~A.}\ \bibnamefont {Witten}},\ }\bibfield  {title} {\bibinfo {title} {Geometric origin of excess low-frequency vibrational modes in amorphous solids},\ }\href {https://doi.org/10.1209/epl/i2005-10245-5} {\bibfield  {journal} {\bibinfo  {journal} {Europhysics Letters (EPL)}\ }\textbf {\bibinfo {volume} {72}},\ \bibinfo {pages} {486} (\bibinfo {year} {2005})},\ \bibinfo {note} {arXiv: cond-mat/0409687}\BibitemShut {NoStop}%
\bibitem [{\citenamefont {Ikeda}\ \emph {et~al.}(2013)\citenamefont {Ikeda}, \citenamefont {Berthier},\ and\ \citenamefont {Biroli}}]{ikeda_dynamic_2013}%
  \BibitemOpen
  \bibfield  {author} {\bibinfo {author} {\bibfnamefont {A.}~\bibnamefont {Ikeda}}, \bibinfo {author} {\bibfnamefont {L.}~\bibnamefont {Berthier}},\ and\ \bibinfo {author} {\bibfnamefont {G.}~\bibnamefont {Biroli}},\ }\bibfield  {title} {\bibinfo {title} {Dynamic criticality at the jamming transition},\ }\href {https://doi.org/10.1063/1.4769251} {\bibfield  {journal} {\bibinfo  {journal} {The Journal of Chemical Physics}\ }\textbf {\bibinfo {volume} {138}},\ \bibinfo {pages} {12A507} (\bibinfo {year} {2013})},\ \bibinfo {note} {arXiv: 1209.2814}\BibitemShut {NoStop}%
\bibitem [{\citenamefont {Lerner}\ \emph {et~al.}(2014)\citenamefont {Lerner}, \citenamefont {DeGiuli}, \citenamefont {Düring},\ and\ \citenamefont {Wyart}}]{lerner_breakdown_2014}%
  \BibitemOpen
  \bibfield  {author} {\bibinfo {author} {\bibfnamefont {E.}~\bibnamefont {Lerner}}, \bibinfo {author} {\bibfnamefont {E.}~\bibnamefont {DeGiuli}}, \bibinfo {author} {\bibfnamefont {G.}~\bibnamefont {Düring}},\ and\ \bibinfo {author} {\bibfnamefont {M.}~\bibnamefont {Wyart}},\ }\bibfield  {title} {\bibinfo {title} {Breakdown of continuum elasticity in amorphous solids},\ }\href {https://doi.org/10.1039/c4sm00311j} {\bibfield  {journal} {\bibinfo  {journal} {Soft Matter}\ }\textbf {\bibinfo {volume} {10}},\ \bibinfo {pages} {5085} (\bibinfo {year} {2014})}\BibitemShut {NoStop}%
\bibitem [{\citenamefont {Mizuno}\ \emph {et~al.}(2016)\citenamefont {Mizuno}, \citenamefont {Silbert},\ and\ \citenamefont {Sperl}}]{mizuno_spatial_2016}%
  \BibitemOpen
  \bibfield  {author} {\bibinfo {author} {\bibfnamefont {H.}~\bibnamefont {Mizuno}}, \bibinfo {author} {\bibfnamefont {L.~E.}\ \bibnamefont {Silbert}},\ and\ \bibinfo {author} {\bibfnamefont {M.}~\bibnamefont {Sperl}},\ }\bibfield  {title} {\bibinfo {title} {Spatial {Distributions} of {Local} {Elastic} {Moduli} {Near} the {Jamming} {Transition}},\ }\href {https://doi.org/10.1103/PhysRevLett.116.068302} {\bibfield  {journal} {\bibinfo  {journal} {Physical Review Letters}\ }\textbf {\bibinfo {volume} {116}},\ \bibinfo {pages} {068302} (\bibinfo {year} {2016})}\BibitemShut {NoStop}%
\bibitem [{\citenamefont {Lerner}\ and\ \citenamefont {Bouchbinder}(2021)}]{lerner_low-energy_2021}%
  \BibitemOpen
  \bibfield  {author} {\bibinfo {author} {\bibfnamefont {E.}~\bibnamefont {Lerner}}\ and\ \bibinfo {author} {\bibfnamefont {E.}~\bibnamefont {Bouchbinder}},\ }\bibfield  {title} {\bibinfo {title} {Low-energy quasilocalized excitations in structural glasses},\ }\href {https://doi.org/10.1063/5.0069477} {\bibfield  {journal} {\bibinfo  {journal} {The Journal of Chemical Physics}\ }\textbf {\bibinfo {volume} {155}},\ \bibinfo {pages} {200901} (\bibinfo {year} {2021})}\BibitemShut {NoStop}%
\bibitem [{\citenamefont {Silbert}\ \emph {et~al.}(2005)\citenamefont {Silbert}, \citenamefont {Liu},\ and\ \citenamefont {Nagel}}]{silbert_vibrations_2005}%
  \BibitemOpen
  \bibfield  {author} {\bibinfo {author} {\bibfnamefont {L.~E.}\ \bibnamefont {Silbert}}, \bibinfo {author} {\bibfnamefont {A.~J.}\ \bibnamefont {Liu}},\ and\ \bibinfo {author} {\bibfnamefont {S.~R.}\ \bibnamefont {Nagel}},\ }\bibfield  {title} {\bibinfo {title} {Vibrations and {Diverging} {Length} {Scales} {Near} the {Unjamming} {Transition}},\ }\bibfield  {journal} {\bibinfo  {journal} {Physical Review Letters}\ }\textbf {\bibinfo {volume} {95}},\ \href {https://doi.org/10.1103/PhysRevLett.95.098301} {10.1103/PhysRevLett.95.098301} (\bibinfo {year} {2005})\BibitemShut {NoStop}%
\bibitem [{\citenamefont {Wyart}(2010)}]{wyart_scaling_2010}%
  \BibitemOpen
  \bibfield  {author} {\bibinfo {author} {\bibfnamefont {M.}~\bibnamefont {Wyart}},\ }\bibfield  {title} {\bibinfo {title} {Scaling of phononic transport with connectivity in amorphous solids},\ }\href {https://doi.org/10.1209/0295-5075/89/64001} {\bibfield  {journal} {\bibinfo  {journal} {Europhysics Letters}\ }\textbf {\bibinfo {volume} {89}},\ \bibinfo {pages} {64001} (\bibinfo {year} {2010})}\BibitemShut {NoStop}%
\bibitem [{\citenamefont {Schoenholz}\ \emph {et~al.}(2013)\citenamefont {Schoenholz}, \citenamefont {Goodrich}, \citenamefont {Kogan}, \citenamefont {Liu},\ and\ \citenamefont {Nagel}}]{schoenholz_stability_2013}%
  \BibitemOpen
  \bibfield  {author} {\bibinfo {author} {\bibfnamefont {S.~S.}\ \bibnamefont {Schoenholz}}, \bibinfo {author} {\bibfnamefont {C.~P.}\ \bibnamefont {Goodrich}}, \bibinfo {author} {\bibfnamefont {O.}~\bibnamefont {Kogan}}, \bibinfo {author} {\bibfnamefont {A.~J.}\ \bibnamefont {Liu}},\ and\ \bibinfo {author} {\bibfnamefont {S.~R.}\ \bibnamefont {Nagel}},\ }\bibfield  {title} {\bibinfo {title} {Stability of jammed packings {II}: the transverse length scale},\ }\href {https://doi.org/10.1039/c3sm51096d} {\bibfield  {journal} {\bibinfo  {journal} {Soft Matter}\ }\textbf {\bibinfo {volume} {9}},\ \bibinfo {pages} {11000} (\bibinfo {year} {2013})}\BibitemShut {NoStop}%
\bibitem [{\citenamefont {Cakir}\ and\ \citenamefont {Pica~Ciamarra}(2016)}]{cakir_emergence_2016}%
  \BibitemOpen
  \bibfield  {author} {\bibinfo {author} {\bibfnamefont {A.}~\bibnamefont {Cakir}}\ and\ \bibinfo {author} {\bibfnamefont {M.}~\bibnamefont {Pica~Ciamarra}},\ }\bibfield  {title} {\bibinfo {title} {Emergence of linear elasticity from the atomistic description of matter},\ }\href {https://doi.org/10.1063/1.4960184} {\bibfield  {journal} {\bibinfo  {journal} {The Journal of Chemical Physics}\ }\textbf {\bibinfo {volume} {145}},\ \bibinfo {pages} {054507} (\bibinfo {year} {2016})}\BibitemShut {NoStop}%
\bibitem [{\citenamefont {Rens}\ and\ \citenamefont {Lerner}(2019)}]{Rens2019}%
  \BibitemOpen
  \bibfield  {author} {\bibinfo {author} {\bibfnamefont {R.}~\bibnamefont {Rens}}\ and\ \bibinfo {author} {\bibfnamefont {E.}~\bibnamefont {Lerner}},\ }\bibfield  {title} {\bibinfo {title} {Rigidity and auxeticity transitions in networks with strong bond-bending interactions},\ }\href {https://doi.org/10.1140/epje/i2019-11888-5} {\bibfield  {journal} {\bibinfo  {journal} {Eur. Phys. J. B}\ }\textbf {\bibinfo {volume} {42}},\ \bibinfo {pages} {114} (\bibinfo {year} {2019})}\BibitemShut {NoStop}%
\bibitem [{\citenamefont {DeGiuli}\ \emph {et~al.}(2014{\natexlab{a}})\citenamefont {DeGiuli}, \citenamefont {Laversanne-Finot}, \citenamefont {D{\"u}ring}, \citenamefont {Lerner},\ and\ \citenamefont {Wyart}}]{degiuli2014effects}%
  \BibitemOpen
  \bibfield  {author} {\bibinfo {author} {\bibfnamefont {E.}~\bibnamefont {DeGiuli}}, \bibinfo {author} {\bibfnamefont {A.}~\bibnamefont {Laversanne-Finot}}, \bibinfo {author} {\bibfnamefont {G.}~\bibnamefont {D{\"u}ring}}, \bibinfo {author} {\bibfnamefont {E.}~\bibnamefont {Lerner}},\ and\ \bibinfo {author} {\bibfnamefont {M.}~\bibnamefont {Wyart}},\ }\bibfield  {title} {\bibinfo {title} {Effects of coordination and pressure on sound attenuation, boson peak and elasticity in amorphous solids},\ }\href@noop {} {\bibfield  {journal} {\bibinfo  {journal} {Soft matter}\ }\textbf {\bibinfo {volume} {10}},\ \bibinfo {pages} {5628} (\bibinfo {year} {2014}{\natexlab{a}})}\BibitemShut {NoStop}%
\bibitem [{\citenamefont {Goodrich}\ \emph {et~al.}(2014)\citenamefont {Goodrich}, \citenamefont {Dagois-Bohy}, \citenamefont {Tighe}, \citenamefont {Van~Hecke}, \citenamefont {Liu},\ and\ \citenamefont {Nagel}}]{goodrich_jamming_2014}%
  \BibitemOpen
  \bibfield  {author} {\bibinfo {author} {\bibfnamefont {C.~P.}\ \bibnamefont {Goodrich}}, \bibinfo {author} {\bibfnamefont {S.}~\bibnamefont {Dagois-Bohy}}, \bibinfo {author} {\bibfnamefont {B.~P.}\ \bibnamefont {Tighe}}, \bibinfo {author} {\bibfnamefont {M.}~\bibnamefont {Van~Hecke}}, \bibinfo {author} {\bibfnamefont {A.~J.}\ \bibnamefont {Liu}},\ and\ \bibinfo {author} {\bibfnamefont {S.~R.}\ \bibnamefont {Nagel}},\ }\bibfield  {title} {\bibinfo {title} {Jamming in finite systems: {Stability}, anisotropy, fluctuations, and scaling},\ }\href {https://doi.org/10.1103/PhysRevE.90.022138} {\bibfield  {journal} {\bibinfo  {journal} {Physical Review E}\ }\textbf {\bibinfo {volume} {90}},\ \bibinfo {pages} {022138} (\bibinfo {year} {2014})}\BibitemShut {NoStop}%
\bibitem [{\citenamefont {Kapteijns}\ \emph {et~al.}(2021{\natexlab{a}})\citenamefont {Kapteijns}, \citenamefont {Bouchbinder},\ and\ \citenamefont {Lerner}}]{kapteijns_unified_2021}%
  \BibitemOpen
  \bibfield  {author} {\bibinfo {author} {\bibfnamefont {G.}~\bibnamefont {Kapteijns}}, \bibinfo {author} {\bibfnamefont {E.}~\bibnamefont {Bouchbinder}},\ and\ \bibinfo {author} {\bibfnamefont {E.}~\bibnamefont {Lerner}},\ }\bibfield  {title} {\bibinfo {title} {Unified quantifier of mechanical disorder in solids},\ }\href {https://doi.org/10.1103/PhysRevE.104.035001} {\bibfield  {journal} {\bibinfo  {journal} {Physical Review E}\ }\textbf {\bibinfo {volume} {104}},\ \bibinfo {pages} {035001} (\bibinfo {year} {2021}{\natexlab{a}})},\ \bibinfo {note} {publisher: American Physical Society}\BibitemShut {NoStop}%
\bibitem [{\citenamefont {González-López}\ \emph {et~al.}(2023)\citenamefont {González-López}, \citenamefont {Bouchbinder},\ and\ \citenamefont {Lerner}}]{gonzalez-lopez_variability_2023}%
  \BibitemOpen
  \bibfield  {author} {\bibinfo {author} {\bibfnamefont {K.}~\bibnamefont {González-López}}, \bibinfo {author} {\bibfnamefont {E.}~\bibnamefont {Bouchbinder}},\ and\ \bibinfo {author} {\bibfnamefont {E.}~\bibnamefont {Lerner}},\ }\bibfield  {title} {\bibinfo {title} {Variability of mesoscopic mechanical disorder in disordered solids},\ }\href {https://doi.org/10.1016/j.jnoncrysol.2023.122137} {\bibfield  {journal} {\bibinfo  {journal} {Journal of Non-Crystalline Solids}\ }\textbf {\bibinfo {volume} {604}},\ \bibinfo {pages} {122137} (\bibinfo {year} {2023})}\BibitemShut {NoStop}%
\bibitem [{\citenamefont {González-López}\ \emph {et~al.}(2021)\citenamefont {González-López}, \citenamefont {Shivam}, \citenamefont {Zheng}, \citenamefont {Ciamarra},\ and\ \citenamefont {Lerner}}]{gonzalez-lopez_mechanical_2021}%
  \BibitemOpen
  \bibfield  {author} {\bibinfo {author} {\bibfnamefont {K.}~\bibnamefont {González-López}}, \bibinfo {author} {\bibfnamefont {M.}~\bibnamefont {Shivam}}, \bibinfo {author} {\bibfnamefont {Y.}~\bibnamefont {Zheng}}, \bibinfo {author} {\bibfnamefont {M.~P.}\ \bibnamefont {Ciamarra}},\ and\ \bibinfo {author} {\bibfnamefont {E.}~\bibnamefont {Lerner}},\ }\bibfield  {title} {\bibinfo {title} {Mechanical disorder of sticky-sphere glasses. {I}. {Effect} of attractive interactions},\ }\href {https://doi.org/10.1103/PhysRevE.103.022605} {\bibfield  {journal} {\bibinfo  {journal} {Physical Review E}\ }\textbf {\bibinfo {volume} {103}},\ \bibinfo {pages} {022605} (\bibinfo {year} {2021})},\ \bibinfo {note} {arXiv: 2008.12011}\BibitemShut {NoStop}%
\bibitem [{\citenamefont {Lerner}\ and\ \citenamefont {Bouchbinder}(2023)}]{lerner_anomalous_2023}%
  \BibitemOpen
  \bibfield  {author} {\bibinfo {author} {\bibfnamefont {E.}~\bibnamefont {Lerner}}\ and\ \bibinfo {author} {\bibfnamefont {E.}~\bibnamefont {Bouchbinder}},\ }\bibfield  {title} {\bibinfo {title} {Anomalous linear elasticity of disordered networks},\ }\href {https://doi.org/10.1039/D2SM01253G} {\bibfield  {journal} {\bibinfo  {journal} {Soft Matter}\ }\textbf {\bibinfo {volume} {19}},\ \bibinfo {pages} {1076} (\bibinfo {year} {2023})},\ \bibinfo {note} {publisher: The Royal Society of Chemistry}\BibitemShut {NoStop}%
\bibitem [{\citenamefont {Wyart}(2005)}]{matthieu_thesis}%
  \BibitemOpen
  \bibfield  {author} {\bibinfo {author} {\bibfnamefont {M.}~\bibnamefont {Wyart}},\ }\bibfield  {title} {\bibinfo {title} {On the rigidity of amorphous solids},\ }\href {https://doi.org/10.1051/anphys:2006003} {\bibfield  {journal} {\bibinfo  {journal} {Ann. Phys. Fr.}\ }\textbf {\bibinfo {volume} {30}},\ \bibinfo {pages} {1} (\bibinfo {year} {2005})}\BibitemShut {NoStop}%
\bibitem [{\citenamefont {Hexner}\ \emph {et~al.}(2019)\citenamefont {Hexner}, \citenamefont {Urbani},\ and\ \citenamefont {Zamponi}}]{hexner_can_2019}%
  \BibitemOpen
  \bibfield  {author} {\bibinfo {author} {\bibfnamefont {D.}~\bibnamefont {Hexner}}, \bibinfo {author} {\bibfnamefont {P.}~\bibnamefont {Urbani}},\ and\ \bibinfo {author} {\bibfnamefont {F.}~\bibnamefont {Zamponi}},\ }\bibfield  {title} {\bibinfo {title} {Can a {Large} {Packing} be {Assembled} from {Smaller} {Ones}?},\ }\href {https://doi.org/10.1103/PhysRevLett.123.068003} {\bibfield  {journal} {\bibinfo  {journal} {Physical Review Letters}\ }\textbf {\bibinfo {volume} {123}},\ \bibinfo {pages} {068003} (\bibinfo {year} {2019})}\BibitemShut {NoStop}%
\bibitem [{\citenamefont {Wang}\ \emph {et~al.}(2019)\citenamefont {Wang}, \citenamefont {Berthier}, \citenamefont {Flenner}, \citenamefont {Guan},\ and\ \citenamefont {Szamel}}]{wang_sound_2019}%
  \BibitemOpen
  \bibfield  {author} {\bibinfo {author} {\bibfnamefont {L.}~\bibnamefont {Wang}}, \bibinfo {author} {\bibfnamefont {L.}~\bibnamefont {Berthier}}, \bibinfo {author} {\bibfnamefont {E.}~\bibnamefont {Flenner}}, \bibinfo {author} {\bibfnamefont {P.}~\bibnamefont {Guan}},\ and\ \bibinfo {author} {\bibfnamefont {G.}~\bibnamefont {Szamel}},\ }\bibfield  {title} {\bibinfo {title} {Sound attenuation in stable glasses},\ }\href {https://doi.org/10.1039/C9SM01092K} {\bibfield  {journal} {\bibinfo  {journal} {Soft Matter}\ }\textbf {\bibinfo {volume} {15}},\ \bibinfo {pages} {7018} (\bibinfo {year} {2019})},\ \bibinfo {note} {publisher: The Royal Society of Chemistry}\BibitemShut {NoStop}%
\bibitem [{\citenamefont {DeGiuli}\ \emph {et~al.}(2014{\natexlab{b}})\citenamefont {DeGiuli}, \citenamefont {Laversanne-Finot}, \citenamefont {Düring}, \citenamefont {Lerner},\ and\ \citenamefont {Wyart}}]{degiuli_effects_2014}%
  \BibitemOpen
  \bibfield  {author} {\bibinfo {author} {\bibfnamefont {E.}~\bibnamefont {DeGiuli}}, \bibinfo {author} {\bibfnamefont {A.}~\bibnamefont {Laversanne-Finot}}, \bibinfo {author} {\bibfnamefont {G.}~\bibnamefont {Düring}}, \bibinfo {author} {\bibfnamefont {E.}~\bibnamefont {Lerner}},\ and\ \bibinfo {author} {\bibfnamefont {M.}~\bibnamefont {Wyart}},\ }\bibfield  {title} {\bibinfo {title} {Effects of coordination and pressure on sound attenuation, boson peak and elasticity in amorphous solids},\ }\href {https://doi.org/10.1039/C4SM00561A} {\bibfield  {journal} {\bibinfo  {journal} {Soft Matter}\ }\textbf {\bibinfo {volume} {10}},\ \bibinfo {pages} {5628} (\bibinfo {year} {2014}{\natexlab{b}})},\ \bibinfo {note} {publisher: The Royal Society of Chemistry}\BibitemShut {NoStop}%
\bibitem [{\citenamefont {Kapteijns}\ \emph {et~al.}(2021{\natexlab{b}})\citenamefont {Kapteijns}, \citenamefont {Richard}, \citenamefont {Bouchbinder},\ and\ \citenamefont {Lerner}}]{kapteijns_elastic_2021}%
  \BibitemOpen
  \bibfield  {author} {\bibinfo {author} {\bibfnamefont {G.}~\bibnamefont {Kapteijns}}, \bibinfo {author} {\bibfnamefont {D.}~\bibnamefont {Richard}}, \bibinfo {author} {\bibfnamefont {E.}~\bibnamefont {Bouchbinder}},\ and\ \bibinfo {author} {\bibfnamefont {E.}~\bibnamefont {Lerner}},\ }\bibfield  {title} {\bibinfo {title} {Elastic moduli fluctuations predict wave attenuation rates in glasses},\ }\href {https://doi.org/10.1063/5.0038710} {\bibfield  {journal} {\bibinfo  {journal} {The Journal of Chemical Physics}\ }\textbf {\bibinfo {volume} {154}},\ \bibinfo {pages} {081101} (\bibinfo {year} {2021}{\natexlab{b}})}\BibitemShut {NoStop}%
\bibitem [{\citenamefont {Szamel}\ and\ \citenamefont {Flenner}(2022)}]{szamel_microscopic_2022}%
  \BibitemOpen
  \bibfield  {author} {\bibinfo {author} {\bibfnamefont {G.}~\bibnamefont {Szamel}}\ and\ \bibinfo {author} {\bibfnamefont {E.}~\bibnamefont {Flenner}},\ }\bibfield  {title} {\bibinfo {title} {Microscopic analysis of sound attenuation in low-temperature amorphous solids reveals quantitative importance of non-affine effects},\ }\href {https://doi.org/10.1063/5.0085199} {\bibfield  {journal} {\bibinfo  {journal} {The Journal of Chemical Physics}\ }\textbf {\bibinfo {volume} {156}},\ \bibinfo {pages} {144502} (\bibinfo {year} {2022})},\ \bibinfo {note} {arXiv:2107.14254 [cond-mat]}\BibitemShut {NoStop}%
\bibitem [{\citenamefont {Lerner}(2019)}]{lerner_mechanical_2019}%
  \BibitemOpen
  \bibfield  {author} {\bibinfo {author} {\bibfnamefont {E.}~\bibnamefont {Lerner}},\ }\bibfield  {title} {\bibinfo {title} {Mechanical properties of simple computer glasses},\ }\href {https://doi.org/10.1016/j.jnoncrysol.2019.119570} {\bibfield  {journal} {\bibinfo  {journal} {Journal of Non-Crystalline Solids}\ }\textbf {\bibinfo {volume} {522}},\ \bibinfo {pages} {119570} (\bibinfo {year} {2019})},\ \bibinfo {note} {arXiv: 1902.08991}\BibitemShut {NoStop}%
\bibitem [{\citenamefont {Bitzek}\ \emph {et~al.}(2006)\citenamefont {Bitzek}, \citenamefont {Koskinen}, \citenamefont {Gähler}, \citenamefont {Moseler},\ and\ \citenamefont {Gumbsch}}]{bitzek_structural_2006}%
  \BibitemOpen
  \bibfield  {author} {\bibinfo {author} {\bibfnamefont {E.}~\bibnamefont {Bitzek}}, \bibinfo {author} {\bibfnamefont {P.}~\bibnamefont {Koskinen}}, \bibinfo {author} {\bibfnamefont {F.}~\bibnamefont {Gähler}}, \bibinfo {author} {\bibfnamefont {M.}~\bibnamefont {Moseler}},\ and\ \bibinfo {author} {\bibfnamefont {P.}~\bibnamefont {Gumbsch}},\ }\bibfield  {title} {\bibinfo {title} {Structural {Relaxation} {Made} {Simple}},\ }\bibfield  {journal} {\bibinfo  {journal} {Physical Review Letters}\ }\textbf {\bibinfo {volume} {97}},\ \href {https://doi.org/10.1103/PhysRevLett.97.170201} {10.1103/PhysRevLett.97.170201} (\bibinfo {year} {2006})\BibitemShut {NoStop}%
\bibitem [{\citenamefont {Lerner}\ and\ \citenamefont {Bouchbinder}(2018)}]{lerner_characteristic_2018}%
  \BibitemOpen
  \bibfield  {author} {\bibinfo {author} {\bibfnamefont {E.}~\bibnamefont {Lerner}}\ and\ \bibinfo {author} {\bibfnamefont {E.}~\bibnamefont {Bouchbinder}},\ }\bibfield  {title} {\bibinfo {title} {A characteristic energy scale in glasses},\ }\href {https://doi.org/10.1063/1.5024776} {\bibfield  {journal} {\bibinfo  {journal} {The Journal of Chemical Physics}\ }\textbf {\bibinfo {volume} {148}},\ \bibinfo {pages} {214502} (\bibinfo {year} {2018})},\ \bibinfo {note} {publisher: American Institute of Physics}\BibitemShut {NoStop}%
\bibitem [{\citenamefont {Hentschel}\ \emph {et~al.}(2011)\citenamefont {Hentschel}, \citenamefont {Karmakar}, \citenamefont {Lerner},\ and\ \citenamefont {Procaccia}}]{hentschel_athermal_2011}%
  \BibitemOpen
  \bibfield  {author} {\bibinfo {author} {\bibfnamefont {H.~G.~E.}\ \bibnamefont {Hentschel}}, \bibinfo {author} {\bibfnamefont {S.}~\bibnamefont {Karmakar}}, \bibinfo {author} {\bibfnamefont {E.}~\bibnamefont {Lerner}},\ and\ \bibinfo {author} {\bibfnamefont {I.}~\bibnamefont {Procaccia}},\ }\bibfield  {title} {\bibinfo {title} {Do athermal amorphous solids exist?},\ }\href {https://doi.org/10.1103/PhysRevE.83.061101} {\bibfield  {journal} {\bibinfo  {journal} {Physical Review E}\ }\textbf {\bibinfo {volume} {83}},\ \bibinfo {pages} {061101} (\bibinfo {year} {2011})}\BibitemShut {NoStop}%
\bibitem [{\citenamefont {Richard}\ \emph {et~al.}(2021)\citenamefont {Richard}, \citenamefont {Lerner},\ and\ \citenamefont {Bouchbinder}}]{richard_brittle--ductile_2021}%
  \BibitemOpen
  \bibfield  {author} {\bibinfo {author} {\bibfnamefont {D.}~\bibnamefont {Richard}}, \bibinfo {author} {\bibfnamefont {E.}~\bibnamefont {Lerner}},\ and\ \bibinfo {author} {\bibfnamefont {E.}~\bibnamefont {Bouchbinder}},\ }\bibfield  {title} {\bibinfo {title} {Brittle-to-ductile transitions in glasses: {Roles} of soft defects and loading geometry},\ }\bibfield  {journal} {\bibinfo  {journal} {MRS Bulletin}\ }\href {https://doi.org/10.1557/s43577-021-00171-8} {10.1557/s43577-021-00171-8} (\bibinfo {year} {2021})\BibitemShut {NoStop}%
\bibitem [{\citenamefont {Schirmacher}\ and\ \citenamefont {Ruocco}(2022)}]{ramos_heterogeneous_2022}%
  \BibitemOpen
  \bibfield  {author} {\bibinfo {author} {\bibfnamefont {W.}~\bibnamefont {Schirmacher}}\ and\ \bibinfo {author} {\bibfnamefont {G.}~\bibnamefont {Ruocco}},\ }\bibfield  {title} {\bibinfo {title} {Heterogeneous {Elasticity}: {The} {Tale} of the {Boson} {Peak}},\ }in\ \href {https://doi.org/10.1142/9781800612587_0009} {\emph {\bibinfo {booktitle} {Low-{Temperature} {Thermal} and {Vibrational} {Properties} of {Disordered} {Solids}}}}\ (\bibinfo  {publisher} {WORLD SCIENTIFIC (EUROPE)},\ \bibinfo {year} {2022})\ pp.\ \bibinfo {pages} {331--373}\BibitemShut {NoStop}%
\bibitem [{\citenamefont {Schirmacher}(2011)}]{schirmacher_comments_2011}%
  \BibitemOpen
  \bibfield  {author} {\bibinfo {author} {\bibfnamefont {W.}~\bibnamefont {Schirmacher}},\ }\bibfield  {title} {\bibinfo {title} {Some comments on fluctuating-elasticity and local oscillator models for anomalous vibrational excitations in glasses},\ }\href {https://doi.org/10.1016/j.jnoncrysol.2010.07.052} {\bibfield  {journal} {\bibinfo  {journal} {Journal of Non-Crystalline Solids}\ }\bibinfo {series} {6th {International} {Discussion} {Meeting} on {Relaxation} in {Complex} {Systems}},\ \textbf {\bibinfo {volume} {357}},\ \bibinfo {pages} {518} (\bibinfo {year} {2011})}\BibitemShut {NoStop}%
\bibitem [{\citenamefont {Schirmacher}\ \emph {et~al.}(2007)\citenamefont {Schirmacher}, \citenamefont {Ruocco},\ and\ \citenamefont {Scopigno}}]{schirmacher_acoustic_2007}%
  \BibitemOpen
  \bibfield  {author} {\bibinfo {author} {\bibfnamefont {W.}~\bibnamefont {Schirmacher}}, \bibinfo {author} {\bibfnamefont {G.}~\bibnamefont {Ruocco}},\ and\ \bibinfo {author} {\bibfnamefont {T.}~\bibnamefont {Scopigno}},\ }\bibfield  {title} {\bibinfo {title} {Acoustic {Attenuation} in {Glasses} and its {Relation} with the {Boson} {Peak}},\ }\href {https://doi.org/10.1103/PhysRevLett.98.025501} {\bibfield  {journal} {\bibinfo  {journal} {Physical Review Letters}\ }\textbf {\bibinfo {volume} {98}},\ \bibinfo {pages} {025501} (\bibinfo {year} {2007})},\ \bibinfo {note} {publisher: American Physical Society}\BibitemShut {NoStop}%
\bibitem [{\citenamefont {Schirmacher}(2006)}]{schirmacher_thermal_2006}%
  \BibitemOpen
  \bibfield  {author} {\bibinfo {author} {\bibfnamefont {W.}~\bibnamefont {Schirmacher}},\ }\bibfield  {title} {\bibinfo {title} {Thermal conductivity of glassy materials and the “boson peak"},\ }\href {https://doi.org/10.1209/epl/i2005-10471-9} {\bibfield  {journal} {\bibinfo  {journal} {Europhysics Letters}\ }\textbf {\bibinfo {volume} {73}},\ \bibinfo {pages} {892} (\bibinfo {year} {2006})},\ \bibinfo {note} {publisher: IOP Publishing}\BibitemShut {NoStop}%
\bibitem [{\citenamefont {Mahajan}\ and\ \citenamefont {Ciamarra}(2021)}]{mahajan_unifying_2021}%
  \BibitemOpen
  \bibfield  {author} {\bibinfo {author} {\bibfnamefont {S.}~\bibnamefont {Mahajan}}\ and\ \bibinfo {author} {\bibfnamefont {M.~P.}\ \bibnamefont {Ciamarra}},\ }\bibfield  {title} {\bibinfo {title} {Unifying description of the vibrational anomalies of amorphous materials},\ }\href {https://doi.org/10.1103/PhysRevLett.127.215504} {\bibfield  {journal} {\bibinfo  {journal} {Physical Review Letters}\ }\textbf {\bibinfo {volume} {127}},\ \bibinfo {pages} {215504} (\bibinfo {year} {2021})},\ \bibinfo {note} {arXiv:2106.04868 [cond-mat]}\BibitemShut {NoStop}%
\bibitem [{\citenamefont {Rainone}\ \emph {et~al.}(2020)\citenamefont {Rainone}, \citenamefont {Bouchbinder},\ and\ \citenamefont {Lerner}}]{rainone_pinching_2020}%
  \BibitemOpen
  \bibfield  {author} {\bibinfo {author} {\bibfnamefont {C.}~\bibnamefont {Rainone}}, \bibinfo {author} {\bibfnamefont {E.}~\bibnamefont {Bouchbinder}},\ and\ \bibinfo {author} {\bibfnamefont {E.}~\bibnamefont {Lerner}},\ }\bibfield  {title} {\bibinfo {title} {Pinching a glass reveals key properties of its soft spots},\ }\href {https://doi.org/10.1073/pnas.1919958117} {\bibfield  {journal} {\bibinfo  {journal} {Proceedings of the National Academy of Sciences}\ }\textbf {\bibinfo {volume} {117}},\ \bibinfo {pages} {5228} (\bibinfo {year} {2020})}\BibitemShut {NoStop}%
\bibitem [{\citenamefont {Giannini}\ \emph {et~al.}(2021)\citenamefont {Giannini}, \citenamefont {Richard}, \citenamefont {Manning},\ and\ \citenamefont {Lerner}}]{giannini_bond-space_2021}%
  \BibitemOpen
  \bibfield  {author} {\bibinfo {author} {\bibfnamefont {J.~A.}\ \bibnamefont {Giannini}}, \bibinfo {author} {\bibfnamefont {D.}~\bibnamefont {Richard}}, \bibinfo {author} {\bibfnamefont {M.~L.}\ \bibnamefont {Manning}},\ and\ \bibinfo {author} {\bibfnamefont {E.}~\bibnamefont {Lerner}},\ }\bibfield  {title} {\bibinfo {title} {Bond-space operator disentangles quasilocalized and phononic modes in structural glasses},\ }\href {https://doi.org/10.1103/PhysRevE.104.044905} {\bibfield  {journal} {\bibinfo  {journal} {Physical Review E}\ }\textbf {\bibinfo {volume} {104}},\ \bibinfo {pages} {044905} (\bibinfo {year} {2021})}\BibitemShut {NoStop}%
\bibitem [{\citenamefont {Shimada}\ \emph {et~al.}(2018)\citenamefont {Shimada}, \citenamefont {Mizuno}, \citenamefont {Wyart},\ and\ \citenamefont {Ikeda}}]{shimada_spatial_2018}%
  \BibitemOpen
  \bibfield  {author} {\bibinfo {author} {\bibfnamefont {M.}~\bibnamefont {Shimada}}, \bibinfo {author} {\bibfnamefont {H.}~\bibnamefont {Mizuno}}, \bibinfo {author} {\bibfnamefont {M.}~\bibnamefont {Wyart}},\ and\ \bibinfo {author} {\bibfnamefont {A.}~\bibnamefont {Ikeda}},\ }\bibfield  {title} {\bibinfo {title} {Spatial structure of quasilocalized vibrations in nearly jammed amorphous solids},\ }\href {https://doi.org/10.1103/PhysRevE.98.060901} {\bibfield  {journal} {\bibinfo  {journal} {Physical Review E}\ }\textbf {\bibinfo {volume} {98}},\ \bibinfo {pages} {060901} (\bibinfo {year} {2018})}\BibitemShut {NoStop}%
\bibitem [{\citenamefont {Yan}\ \emph {et~al.}(2016)\citenamefont {Yan}, \citenamefont {DeGiuli},\ and\ \citenamefont {Wyart}}]{yan_variational_2016}%
  \BibitemOpen
  \bibfield  {author} {\bibinfo {author} {\bibfnamefont {L.}~\bibnamefont {Yan}}, \bibinfo {author} {\bibfnamefont {E.}~\bibnamefont {DeGiuli}},\ and\ \bibinfo {author} {\bibfnamefont {M.}~\bibnamefont {Wyart}},\ }\bibfield  {title} {\bibinfo {title} {On variational arguments for vibrational modes near jamming},\ }\href {https://doi.org/10.1209/0295-5075/114/26003} {\bibfield  {journal} {\bibinfo  {journal} {EPL (Europhysics Letters)}\ }\textbf {\bibinfo {volume} {114}},\ \bibinfo {pages} {26003} (\bibinfo {year} {2016})}\BibitemShut {NoStop}%
\bibitem [{\citenamefont {Ji}\ \emph {et~al.}(2022)\citenamefont {Ji}, \citenamefont {de~Geus}, \citenamefont {Agoritsas},\ and\ \citenamefont {Wyart}}]{ji_mean-field_2022}%
  \BibitemOpen
  \bibfield  {author} {\bibinfo {author} {\bibfnamefont {W.}~\bibnamefont {Ji}}, \bibinfo {author} {\bibfnamefont {T.~W.~J.}\ \bibnamefont {de~Geus}}, \bibinfo {author} {\bibfnamefont {E.}~\bibnamefont {Agoritsas}},\ and\ \bibinfo {author} {\bibfnamefont {M.}~\bibnamefont {Wyart}},\ }\bibfield  {title} {\bibinfo {title} {Mean-field description for the architecture of low-energy excitations in glasses},\ }\href {https://doi.org/10.1103/PhysRevE.105.044601} {\bibfield  {journal} {\bibinfo  {journal} {Physical Review E}\ }\textbf {\bibinfo {volume} {105}},\ \bibinfo {pages} {044601} (\bibinfo {year} {2022})},\ \bibinfo {note} {publisher: American Physical Society}\BibitemShut {NoStop}%
\end{thebibliography}%

\end{document}